\documentclass[sigconf, 10pt]{acmart}

\usepackage{blindtext}
\usepackage[utf8]{inputenc}
\usepackage{epsfig,xspace,url}
\usepackage{xcolor}
\usepackage{subcaption}
\usepackage{amsmath}
\usepackage{mathtools}
\usepackage{amsthm}
\usepackage{tikz}
\usepackage{caption}
\usepackage{graphicx}
\usepackage[ruled, lined, linesnumbered, commentsnumbered, longend]{algorithm2e}
\usepackage[noend]{algpseudocode}
\usepackage{url}
\usepackage{colortbl}
\usepackage{thmtools}
\usepackage{thm-restate}

\usepackage{wrapfig}
\usepackage[subtle]{savetrees}
\usepackage{makecell}
\renewcommand\footnotetextcopyrightpermission[1]{} 
\newcommand{\vk}[1]{}
\newcommand{\pg}[1]{}
\newcommand{\im}[1]{}
\newcommand{\pgv}[1]{}

\input{macros}

\let\ACMmaketitle=\maketitle
\renewcommand{\maketitle}{\begingroup\let\footnote=\thanks \ACMmaketitle\endgroup}
\let\height\relax

\begin{document}

\newcommand{\name}{\textsc{Ethereal}\xspace}

\title{Ethereal: Divide and Conquer Network Load Balancing in Large-Scale Distributed Training}

\author{Vamsi Addanki}
\affiliation{
  \institution{TU Berlin}
}

\author{Prateesh Goyal}
\affiliation{%
  \institution{Microsoft Research}
}

\author{Ilias Marinos}
\affiliation{%
  \institution{NVIDIA}%
}

\author{Stefan Schmid}
\affiliation{%
  \institution{TU Berlin}%
}

\renewcommand{\shortauthors}{Addanki et al.}

\sloppy

\begin{abstract}
Large-scale distributed training in production datacenters constitutes a challenging workload bottlenecked by network communication. In response, both major industry players (e.g., Ultra Ethernet Consortium) and parts of academia have surprisingly, and almost unanimously, agreed that packet spraying is \emph{necessary} to improve the performance of large-scale distributed training workloads.

In this paper, we challenge this prevailing belief and pose the question: \emph{How close can singlepath transport come to matching the performance of packet spraying?} We demonstrate that singlepath transport (from a NIC's perspective) is sufficient and can perform nearly as well as ideal packet spraying, particularly in the context of distributed training in CLOS-based topologies. Our assertion is based on four key observations about workloads driven by collective communication patterns: \emph{(i)} flow sizes are known upon arrival, \emph{(ii)} flow sizes are equal within each step of a collective, \emph{(iii)} the completion time of a collective is more critical than individual flow completion times, and \emph{(iv)} flows can be \emph{split} upon arrival to control load balancing directly from the application layer.

We present \name, a simple distributed load balancing algorithm that opportunistically splits flows and assigns paths to each flow in a transparent manner, requiring little to no changes to existing RDMA NICs. 
Our evaluation, spanning a wide range of collective communication algorithms and GPT models using Astra-Sim, shows that \name significantly reduces the completion times by up to $30\%$ compared to packet spraying and by up to $40\%$ compared to REPS, even under link failures. This paper offers an alternative perspective for developing next-generation transport protocols tailored to large-scale distributed training.
 
\end{abstract}

\maketitle
\thispagestyle{plain}
\pagestyle{plain}

\section{Introduction}
\label{sec:introduction}
The rapid increase in the computational needs of emerging deep learning models (\eg GPT-4~\cite{gpt4}) necessitates that the training process be distributed across a large-scale GPU cluster (\eg thousands of GPUs) in a datacenter. To the surprise of datacenter operators and researchers, communication has turned out to be a major bottleneck in distributed training~\cite{arzani2023rethinking}. To speed up the training process, significant research efforts have recently been made, including parallelism strategies~\cite{shoeybi2019megatron,10.1145/3341301.3359646,10.1145/3458817.3476209,295549}, topology engineering~\cite{285119,zhao2022efficient}, and collective optimizers~\cite{nccl,285084,arzani2023rethinking}.
Unsurprisingly, congestion control~\cite{278346,276958,10.1145/2785956.2787510,10.1145/3544216.3544235,10.1145/3387514.3406591,10.1145/3341302.3342085,10.1145/1851182.1851192} and load-balancing~\cite{10.1145/2619239.2626316,259355,10.1145/3098822.3098839,10.1145/2890955.2890968} are the main root causes of communication bottlenecks, two of the most widely studied problems in the literature, especially in the context of datacenters.

\begin{figure}[t]
\centering
\includegraphics[width=1\linewidth]{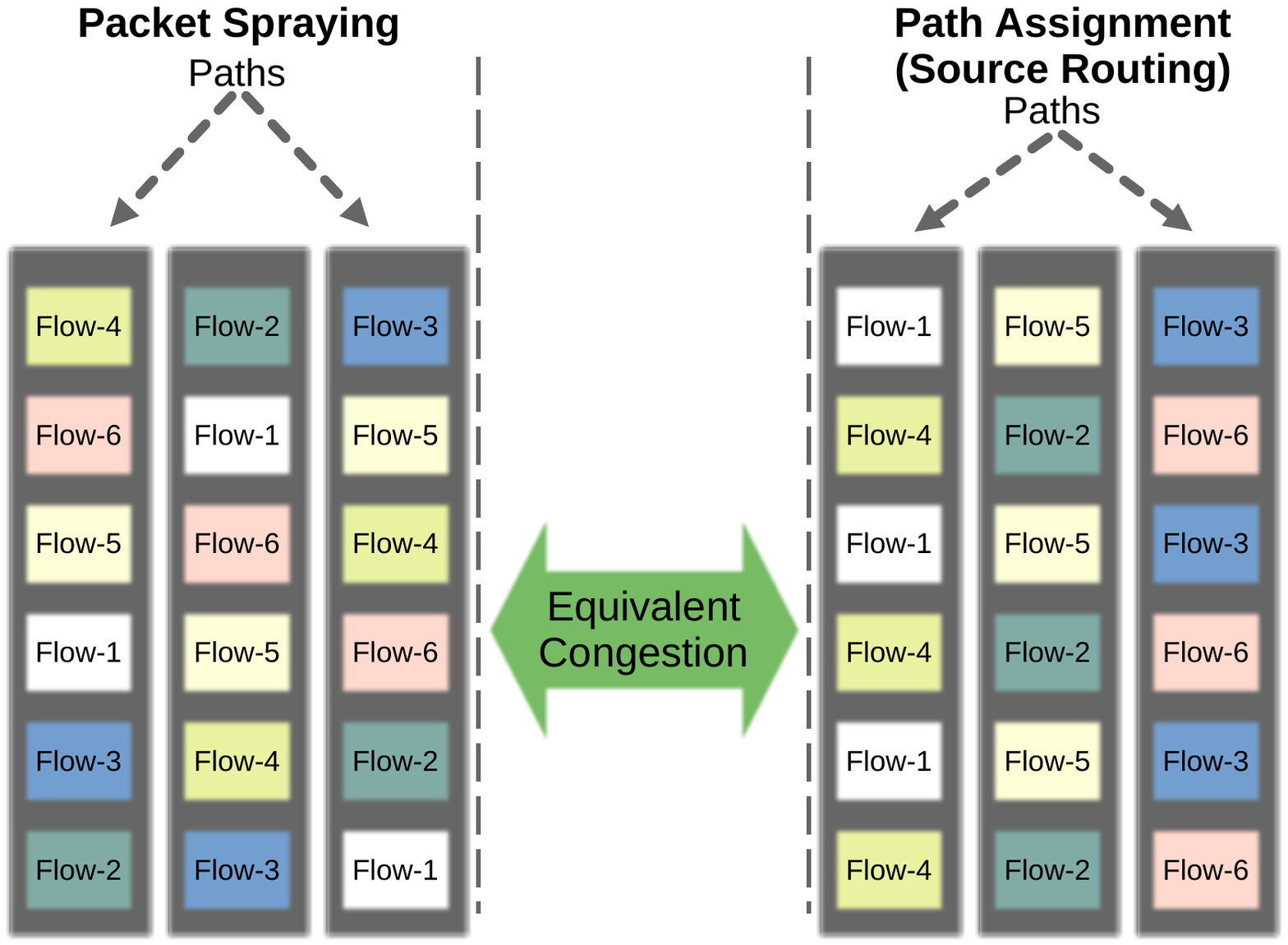}
\caption{\small In contrast to traditional datacenter workloads, distributed training workloads exhibit certain properties in terms of flow sizes, the number of concurrent flows, and arrival times that allow singlepath transport to achieve nearly the same performance as an optimal multipath transport. The problem essentially boils down to assigning paths to each flow in order to minimize congestion.}
\vspace{-4mm}
\label{fig:intro}
\end{figure}

In the wake of this emerging problem, specifically in the context of distributed training, hyperscale datacenter operators have explored several approaches in the recent past --- starting from pure single path load balancing directly at the application layer by Alibaba~\cite{10.1145/3651890.3672265}, splitting every flow into multiple subflows to increase ECMP entropy by Meta~\cite{10.1145/3651890.3672233}, all the way to complex new NIC designs with packet spraying, retransmit buffers and loss recovery mechanisms being explored by major hyperscalers within the Ultra Ethernet Consortium~\cite{uec}.

To our surprise, both industry and academia have largely come to the conclusion that multipath transport with packet spraying is \emph{necessary} to improve the performance of large-scale distributed training workloads~\cite{uec,bonato2024smartt,reps2024,10.1145/3651890.3672233,295627}. Multipath transport protocols are motivated by their ability to exploit path diversity and effectively balance the load across all network paths~\cite{10.1145/3098822.3098825,10.1145/2578901,10.1007/978-3-642-20757-0_35,266964, eqds, homa}.
While, in theory, multipath transport protocols can achieve similar objectives to an optimal multi-commodity flow problem, several practical challenges remain. For instance, packet ordering (and the required reorder buffers), loss recovery, and telemetry (for traffic analysis) complicate NIC design and operational costs. Further, such a complex NIC design in proprietary space often results in vendor lock-in scenarios which is of major concern to hyperscalers. This begs the question:

\medskip
\noindent \textit{Is packet spraying necessary for improving large-scale distributed training performance, and how close can singlepath transport get to the optimal performance?}
\medskip

Diminishing hopes for singlepath transport, it is well known that achieving ideal load balancing is not feasible for an arbitrary set of flows in the network~\cite{7588075}. Moreover, link failures are a common concern in large-scale datacenters~\cite{reps2024,uec}, particularly in collective communication, where the performance of a single flow can significantly impact the completion time of the entire collective.

\textbf{Yet}, our findings indicate a likely \emph{no} i.e., a singlepath transport with application-layer load balancing can perform nearly as well as a multipath transport with packet spraying, even under failures. While transport protocols leveraging packet spraying are ideally superior to singlepath variants for general workloads, certain properties of training workloads allow us to demonstrate that packet spraying is not a necessity to improve performance.
Our assertion is based on four key properties of distributed training workloads --- driven by collective communication --- that distinguish them from the stochastic flow arrivals typical in traditional datacenter workloads.

First, flow sizes are known upon arrival. Specifically, collective communication operates on fixed-size data chunks, and the size of each chunk is determined at the time of flow creation. Alibaba's HPN already leverages this property for load balancing~\cite{10.1145/3651890.3672265}, allowing servers to accurately track the total outstanding bytes on each path and make informed load-balancing decisions without the need for packet spraying across all available paths.

Second, flow sizes remain equal within each step of a collective operation. For example, in a reduce-scatter operation using the recursive doubling algorithm, the flow size in step $i$ is $2^{-i}$ times the total message size, and this size remains identical across all GPUs participating in the collective. This enables distributed load balancing at the source without requiring a centralized view of the network.

Third, individual flow completion times are less critical; instead, the overall completion time of a collective is paramount, as it directly affects the training duration. Notably, under link failures, singlepath transport can still achieve near-optimal collective completion times since identifying and avoiding a failed path is relatively straightforward --- unlike packet spraying --- making it inherently more resilient to failures.

Fourth, the collective communication library has full control over the number and size of flows at any given time. For instance, NCCL allows users to configure \texttt{NCCL\_IB\_QPS\_PER\_CONNECTION} and \texttt{NCCL\_IB\_SPLIT\_DATA\_ON\_QPS}, which distribute data transmissions between a GPU pair across multiple queue pairs uniformly~\cite{envnccl}. Meta leverages this property to effectively increase entropy for ECMP~\cite{10.1145/3651890.3672233}, without requiring any involvement from the NIC.

Given these properties, we claim that uniformly distributing load across multiple equal-cost paths in the network --- similar to packet spraying --- essentially reduces to assigning paths to each flow in a collective directly at the application layer, within the communication library. While prior work has leveraged these properties to some extent, our study is the first to systematically explore their implications for transport protocol design in distributed training workloads.

To strengthen this observation, we analytically prove that singlepath transport (from a NIC's perspective) with uniform path assignment is equivalent to optimal transport with packet spraying in terms of congestion under collective communication workloads. Our proof relies on the ability to split flows at the application layer, and we show that the required splitting is minimal. Furthermore, to motivate the need for a certain degree of flow splitting, we show that any transport relying on adaptive re-routing without flow splitting --- e.g., REPS~\cite{reps2024} --- cannot generally avoid congestion in the network core and is therefore not optimal for distributed training workloads.

We develop a novel load-balancing algorithm, \name, that greedily assigns paths to each flow such that the load is uniformly spread across each path in the network. \name intercepts flows at the communication library and opportunistically splits flows to achieve near-optimal load balancing. Upon failures, \name adopts a simple re-routing strategy based on a single timeout per queue pair at the NIC. Importantly, \name does not require any changes to NIC hardware and can be implemented entirely in software.

Our evaluations, based on Astra-Sim~\cite{won2023astrasim2,rashidi2020astrasim,astrasimweb} and MLCommons Chakra~\cite{chakra}, show that \name significantly improves completion times for collective communication compared to state-of-the-art approaches. Specifically, \name reduces completion time by up to $30\%$ compared to packet spraying (at the end host) and by up to $40\%$ compared to REPS~\cite{reps2024} and ECMP, demonstrating the effectiveness of singlepath transport for distributed training workloads. Even under failures, \name improves completion times by $68\%$ and $58\%$ on average compared to packet spraying and REPS, respectively. Our results challenge the prevailing belief that packet spraying is \emph{necessary} for improving the performance of distributed training workloads and offer an alternative perspective for developing next-generation transport protocols tailored to large-scale distributed training.

\smallskip
\textit{In the remainder of this paper, our distinction between singlepath and multipath is solely from the network interface card (NIC) flow state and congestion control point-of-view.}

\smallskip
\textit{\textbf{This work does not raise any ethical issues.}}

\section{Motivation}
\label{sec:motivation}

We provide a brief background and motivation for our work, highlighting the limitations of existing approaches to load balancing in GPU clusters. We focus specifically on tree-based topologies, such as leaf-spine, fat-tree, which are commonly used in hyperscale datacenters for distributed training workloads. We first describe the key ingredients for load balancing (\S\ref{sec:ingredients}), followed by a discussion on the limitations of existing approaches (\S\ref{sec:limitations}). We then formally present our main observation that singlepath transport can achieve near-optimal performance compared to packet spraying in distributed training workloads (\S\ref{sec:single-multi}). We conclude this section with a brief discussion on the implications of our findings for network load balancing in large-scale distributed training workloads (\S\ref{sec:design-goals}).

\subsection{Design Choices for Load Balancing}
\label{sec:ingredients}
We dissect the network load balancing problem into three key components: \first granularity of control, \second path selection, and \third re-routing, which are the fundamental knobs any algorithm operates on. Various algorithms in the literature make design choices across the spread of these components, leading to different performance outcomes. Table~\ref{tab:design-space} summarizes the design choices to illustrate the trade-offs in load balancing algorithms.

\myitem{Granularity of control:}  
Upon a flow\footnote{A ``flow'' corresponds to a queue pair between two end-hosts that exchange data. The flow terminates upon receiving completion signals (CQEs) for all work units (WQEs) posted by the flow.} arrival, there are three natural choices. 
First, the entire flow can be transmitted over a single path, which we refer to as \emph{singlepath transport}. Second, the flow can be split into multiple subflows and transmitted over multiple paths—yet still considered \emph{singlepath transport} since the NIC maintains separate queue pairs for each subflow. Third, at the finest granularity, each packet can be assigned a different path in the network, which we refer to as \emph{multipath transport} since the NIC maintains a single queue pair but distributes packets across multiple paths.  

The choice of granularity of control directly impacts the complexity of congestion control algorithms and NIC design. At one end of the spectrum, singlepath transport with flow-level load balancing is simple and requires no significant modifications to NIC design. At the other extreme, multipath transport with per-packet load balancing (\eg packet spraying) necessitates a complex NIC design with reordering buffers, loss recovery mechanisms, and sophisticated telemetry to track flows and analyze traffic patterns.

\myitem{Path selection:} A load-balancing algorithm primarily focuses on path assignment for each unit of control to achieve uniform load distribution. In the context of distributed training workloads, path selection can be handled at the application layer~\cite{10.1145/3651890.3672265}, where the communication library is responsible for launching flows. 

Traditionally, flow sizes were unknown upon arrival, forcing algorithms to make online path assignment decisions based on observed traffic patterns (\eg Hedera~\cite{259355}).  
For instance, consider $8$ flows --- $4$ short and $4$ long --- spread across $2$ available paths. Without prior knowledge of flow sizes, load-balancing algorithms struggle even in such a simple scenario. This unpredictability motivated the development of packet spraying techniques to achieve near-optimal load balancing (\eg NDP~\cite{10.1145/3098822.3098825}, DRILL~\cite{10.1145/3098822.3098839}), greatly simplifying path selection.

\begin{table}[t]
    \centering
    \renewcommand{\arraystretch}{3}
    \setlength{\tabcolsep}{4pt}
    \arrayrulecolor{gray!50}
    \resizebox{\columnwidth}{!}{
    \begin{tabular}{|p{2.4cm}|p{2.6cm}|p{2.8cm}|p{2.6cm}|}
        \hline
        \rowcolor{gray!20}
        \makecell{\textbf{Algorithm}} & \makecell{\textbf{Granularity}} & \makecell{\textbf{Path Selection}} & \makecell{\textbf{Re-routing}} \\
        \hline
        \cellcolor{gray!5} \makecell{Vanilla RDMA} & \cellcolor{green!10} \makecell{Flow-level} & \cellcolor{red!10} \makecell{ECMP} & \cellcolor{red!10} \makecell{None} \\
        \hline
        \cellcolor{gray!5} \makecell{MP-RDMA} & \cellcolor{orange!10} \makecell{Sub-flow} & \cellcolor{red!10} \makecell{ECMP} & \cellcolor{red!10} \makecell{None} \\
        \hline
        \cellcolor{gray!5} \makecell{Packet Spraying\\} & \cellcolor{red!10} \makecell{Per-packet\\} & \cellcolor{green!10} \makecell{Uniform\\Random} & \cellcolor{red!10} \makecell{None\\} \\
        \hline
        \cellcolor{gray!5} \makecell{REPS} & \cellcolor{orange!10} \makecell{Per-Packet \\$+$\\ Per-flow} & \cellcolor{orange!10} \makecell{Adaptive} & \cellcolor{orange!10} \makecell{Reactive} \\
        \hline
        \cellcolor{gray!20} \makecell{\name\\(This work)} & \cellcolor{orange!10} \makecell{Sub-flow} & \cellcolor{green!10} \makecell{Static\\Topology-aware} & \cellcolor{green!10} \makecell{Reactive \\$+$\\ Proactive} \\
        \hline
    \end{tabular}
    }
    \vspace*{2mm}
    \caption{Design choices of load balancing approaches}
    \label{tab:design-space}
    \vspace*{-8mm}
    \label{tab:design-choices}
\end{table}

However, in distributed training workloads, flow sizes are known upon arrival, enabling more informed path selection. In the previous example, if flow sizes are known in advance, assigning $2$ short and $2$ long flows to each path trivially achieves uniform load balancing across the $2$ paths --- matching the benefits of packet spraying while maintaining the simplicity of singlepath transport.

\myitem{Re-routing:}
Not only is path selection important, but the ability to re-route flows (i.e., reassign paths) upon failures is crucial for maintaining performance in the presence of link failures.  
In distributed training workloads, flows on a failed path can delay the completion of a collective, severely impacting overall training duration. Unfortunately, network switches are too slow to detect failures and update routing tables. For instance, a switch takes approximately $100$ms to detect a link failure, during which it continues transmitting packets to the failed port --- leading to excessive packet drops. In contrast, an allReduce operation with a $256$MB message size (typical in GPT models) across $256$ GPUs completes in about $10$ms with $400$Gbps port bandwidth. This implies that collective completion time can increase by an order of magnitude if the load-balancing algorithm does not re-route flows in a timely manner.  

At one extreme, load balancing with singlepath transport can readily detect failures from a ``flow'' context by maintaining a single timer and reacting to timeouts. At the other extreme, multipath transport with packet spraying, while optimal in a symmetric topology, requires complex failure detection mechanisms, potentially involving per-packet timers.

\subsection{Limitations of Existing Approaches}
\label{sec:limitations}

In our discussions with hyperscalers, two main arguments frequently emerged: \first packet spraying is believed to provide uniform load balancing under asymmetries, and \second adaptive routing is considered essential for avoiding congestion. Taking a closer look at these arguments, we identify key challenges that motivate alternative approaches.  

Our focus is on load balancing algorithms that operate at the end-host in a distributed manner, without requiring specialized hardware support from network switches. In the context of distributed training workloads, we examine the trade-offs of packet spraying and REPS --- an adaptive load-balancing algorithm currently being explored by UEC~\cite{uec} --- and highlight opportunities to improve load balancing.

\begin{figure}[t]
\centering
\includegraphics[width=0.8\linewidth]{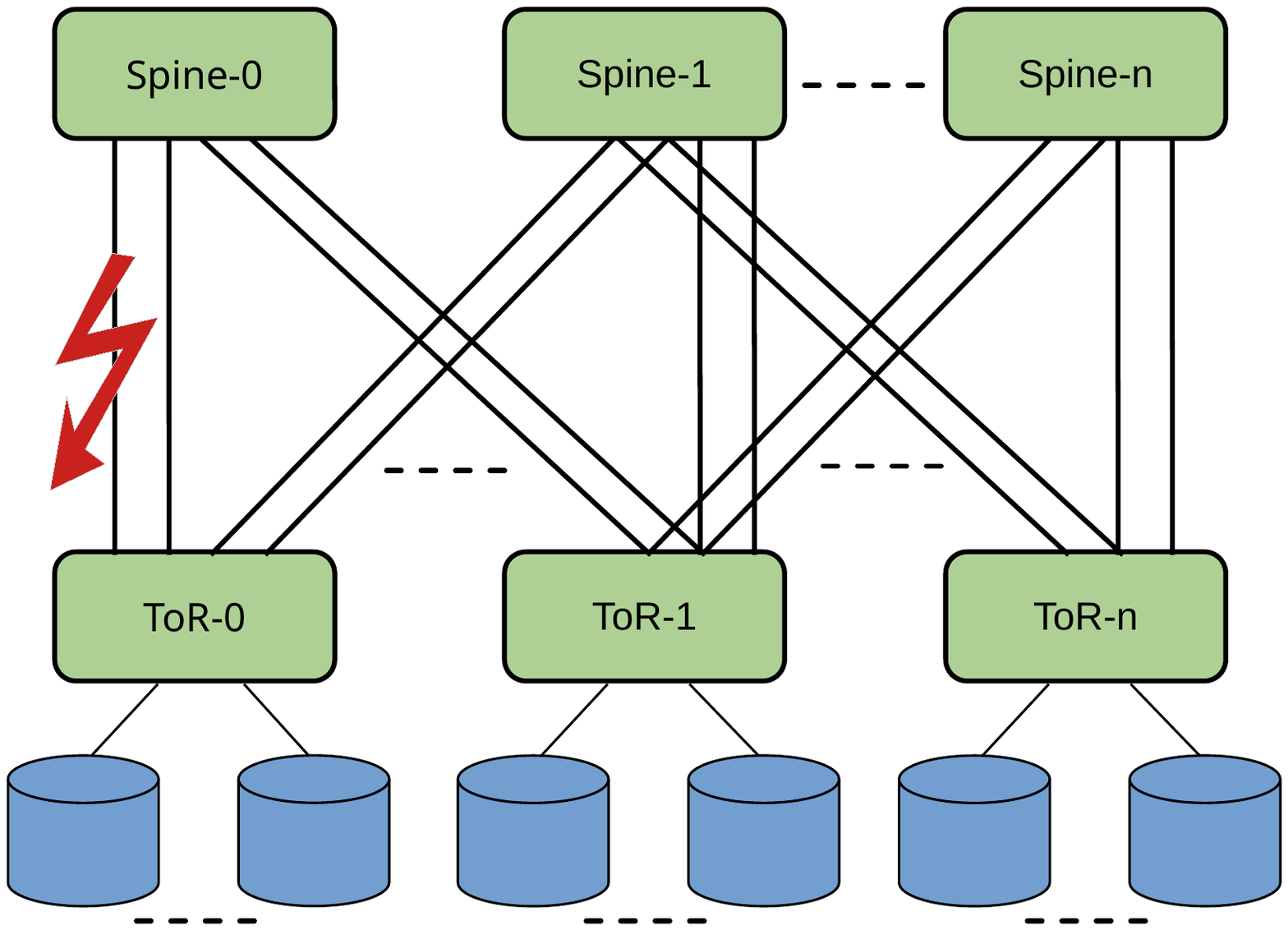}
\caption{Packet spraying struggles with congestion control under asymmetries, whereas, singlepath adaptive load balancing struggles with path flapping even in a symmetric topology.}
\label{fig:limitations-spray}
\end{figure}

\myitem{Severe congestion with packet spraying under failures:} Consider a large leaf-spine topology with $n$ ToRs, $n$ spines, and $2$ links between each ToR and spine switch. Suppose a single link fails between ToR-$0$ and spine-$1$, introducing asymmetry into the topology. Despite this failure, all ToR-spine pairs remain directly connected. Specifically, there are still $2$ paths between ToR-$1$ and spine-$0$, but they are no longer edge-disjoint, now sharing a common bottleneck link.
Packet spraying, which uniformly distributes packets across all available paths, can lead to severe congestion at the slow link. Congestion control struggles to mitigate this issue, as feedback is limited to only two congested paths per round trip --- compared to the $2\cdot n$ paths over which packets are sprayed. This imbalance makes it difficult for congestion control mechanisms to respond effectively, exacerbating congestion in the network core, especially under topological asymmetries.  
In contrast, singlepath transport retains per-path context, allowing for more precise congestion control. By identifying affected paths, impacted flows can adjust their rates accordingly, alleviating congestion and minimizing queuing delays.

\medskip
\noindent{\textcolor{takeawaycolor}{$\blacksquare$ \textbf{\textit{Takeaway.}}} Congestion control with packet spraying is non-trivial under topological asymmetries, leading to severe congestion at slow links. Maintaining per-path context can alleviate congestion by identifying affected paths and adjusting rates accordingly.

\medskip
\myitem{Challenges in avoiding load imbalance with REPS~\cite{reps2024}:}  
REPS is an adaptive load-balancing approach that leverages ECN signals to dynamically re-route flows to less congested paths. This mechanism is particularly effective in rapidly re-routing flows from failed paths, significantly reducing reaction time compared to the $\approx 100$ms required for network switches to detect failures and update ECMP.
While REPS offers key advantages, any adaptive load-balancing algorithm without flow splitting cannot generally avoid congestion in the network core~\cite{7588075}. REPS relies on a reactive re-routing strategy, where flows are redirected only upon receiving an ECN-marked acknowledgment. In certain cases, this approach can converge to a stable state, achieving a near-uniform distribution of flows across paths. However, the total number of flows inherently limits its ability to achieve perfect load balancing.  
For instance, consider $5$ flows from ToR-$1$ to ToR-$n$ in the simple network topology shown in Figure~\ref{fig:limitations-spray} (without any link failures). There are $4$ equal-cost paths but $5$ flows to distribute. Since the number of flows is not evenly divisible by the number of paths, at least two flows must inevitably share a single path. In such cases, REPS waits for queue build-up before re-routing a congested flow upon receiving an ECN signal. However, this reactive strategy can unintentionally cause path flapping, where flows repeatedly switch between paths, leading to persistent congestion in the network core.

\medskip
\noindent{\textcolor{takeawaycolor}{$\blacksquare$ \textbf{\textit{Takeaway.}}} Number of flows fundamentally limits the ability of adaptive load-balancing algorithms to achieve perfect load balancing. 
Adaptive re-routing strategies can lead to path flapping, causing persistent congestion in the network core.

\subsection{Making a Case for Singlepath Transport}
\label{sec:single-multi}

In this section, we seek to understand whether a singlepath\footnote{As discussed in \S\ref{sec:ingredients}, we refer to both per-flow and sub-flow granularity of load balancing as singlepath since they retain the per-path context for congestion control at the NIC.} (per-flow or sub-flow) load-balancing algorithm can achieve nearly the same objective as a multipath (packet spraying) load balancing algorithm.
Specifically, we are interested in simple, distributed algorithms that operate on each end-host without relying on a centralized controller. In the following, we formally establish an equivalence between singlepath load balancing and packet spraying under certain traffic patterns.

\begin{theorem}[Equivalence]\label{th:equiv}
Given a leaf-spine topology, with $\ell$ leaves, $s$ spines and $k$ server nodes, a set of demands $\mathcal{M}=\{ f_{i}\times n_{i,j} \mid f_{i}, n_{i,j}\in \mathbb{N}, i\in [1,k], j\in[1,\ell] \}$, where $f_{i}$ is the flow size and $n_{i,j}$ is the number of flows that a server node $i$ has towards any set of destination nodes within leaf $j$, then a greedy distributed algorithm (at each node) that splits the least number of flows and assigns each flow to the least congested (local perspective) uplink (leaf$\leftrightarrow$spine), is equivalent to packet spraying in terms of the objective to minimize the maximum congestion.
\end{theorem}

We make three key observations about the communication patterns of distributed training that put our result in Theorem~\ref{th:equiv} into context for load-balancing algorithms. We present the formal proof at the end of this section.

\myitem{A single flow can be split to multiple flows:}
While multipath transport protocols split the flow only logically, from a transport point of view, in the context of distributed training, the communication library (traffic source) allows certain user configurations such that each flow can be split into multiple flows with corresponding queue pairs (separate connections) at the NIC. For instance, environment variables in NCCL, such as \texttt{NCCL\_IB\_QPS\_PER\_CONNECTION}, enable splitting the data uniformly into multiple queue pairs. As a result, the number of flows can be controlled in a much more fine-grained manner. We use this property later in the proof of Theorem~\ref{th:equiv} to show that splitting (if needed) only a few number of flows from each source is sufficient to achieve the same properties of an optimal load-balancing algorithm in terms of the maximum congestion.

\myitem{GPU flows are equal in size within each collective step:}
Flows originating from a GPU corresponding to each collective step are of equal size eg., due to chunking, and padding used by NCCL. As result, each $i^{th}$ GPU may generate any number of flows each of size $f_i$ at a given time instance, that fits within the definition of the demands considered in Theorem~\ref{th:equiv}.

\myitem{Fairness across flows is not a requirement:}
Given that our primary concern is the completion time of a collective, fairness across flows is not necessary. Scheduling the flows in any particular order while maintaining uniform load across the network core and high link utilization is critical to achieve fast collective completion times.

Given the workload properties, our result in Theorem~\ref{th:equiv} indicates that a singlepath transport is sufficient for large-scale distributed training. We now present a formal proof.

\begin{proof}[Proof of Theorem~\ref{th:equiv}]
We begin with node $i=1$ and account for the total amount of demand assigned to each uplink and downlink\footnote{We refer to leaf$\rightarrow$spine links as uplinks and spine$\rightarrow$leaf links as downlinks.}. For simplicity, we assume that all demands exit the leaves (use uplinks), focusing on the congestion on both the uplinks and downlinks. There are $n_{1,j}$ flows, each of size $f_1$, towards each destination leaf $j$.
Our greedy algorithm (depicted in Algorithm~\ref{alg:ethereal}), say ALG, operates in steps for each destination leaf $j$. By fixing the destination leaf $j$, choosing an uplink from the source's perspective also implies selecting the corresponding downlink (via the chosen leaf$\rightarrow$spine uplink). Consequently, the demand assigned to each uplink from source $i$ equals the demand assigned to the corresponding downlink.
ALG first assigns $\lfloor\frac{n_{1,j}}{s}\rfloor$ flows to each uplink. At this point, there are $r=n_{1,j}\bmod s$ flows left to be assigned. Let $g = \mathrm{gcd}(r,s)$\footnote{$\mathrm{gcd}(r,s)$ is the greatest common divisor of $r$ and $s$.}. 
\pgv{Writing this math in terms of lcm might be a bit more intuitive.}
ALG splits each of the $r$ flows into $\frac{s}{g}$ flows, each of size $\frac{f_1\cdot g}{s}$, and assigns $\frac{r}{g}$ (an integer) flows of size $\frac{f_1\cdot g}{s}$ to each uplink. In total, ALG assigns $f_1 \cdot \lfloor\frac{n_{1,j}}{s}\rfloor + \frac{f_1\cdot g}{s} \cdot \frac{r}{g} = f_1 \cdot (\lfloor\frac{n_{1,j}}{s}\rfloor + \frac{r}{s}) = f_1 \cdot \frac{n_{i,j}}{s}$ of demand to each uplink.
An optimal multipath load-balancing algorithm (OPT) splits the total demand evenly across all uplinks, assigning $f_1\cdot \frac{n_{i,j}}{s}$ of demand to each uplink. This demonstrates that ALG and OPT assign equal demand to each uplink (and consequently to each downlink) from each source $i$, and the proof follows.
Additionally, at each source, ALG only splits $r=n_{1,j}\bmod s$ flows, and these are split into $\frac{s}{g}$ flows. The extra flows created by ALG at each source are limited to $\frac{r\cdot (s-g)}{g}$.
To prove that the splitting is minimal, assume a better splitting is feasible, with each of the $r$ remaining flows split into $\gamma < \frac{s}{g}$ flows. Thus, the size of each remainder flow is $\frac{f_1}{\gamma}$. We need to assign $r\cdot \gamma$ flows to $s$ uplinks such that the assignment results in $\frac{f_1\cdot r}{s}$ demand on each uplink to be optimal in terms of congestion. This requires $\gamma \ge 1$ to be an integer (for splitting) and $\frac{r\cdot \gamma}{s}$ to be an integer (for assignment). The smallest value of $\gamma$ that satisfies these conditions is $\frac{s}{g}$, contradicting our initial assumption.
\end{proof}

We emphasize that the same result cannot be derived for hash-based ECMP since it does not explicitly converge based on the number of flows, but rather on the entropy of the input to the hash function, which is a non-trivial task to control in practice. Furthermore, such properties cannot be derived for singlepath transport in general. The problem of finding the minimum congestion unsplittable flow is known to be NP-hard~\cite{7588075}. The best-known result for CLOS topologies establishes a tight bound of $4$-approximation for a greedy algorithm~\cite{7588075}. However, the specific properties of collective communication workloads allow us to establish the above results. 

\medskip
\noindent{\textcolor{takeawaycolor}{$\blacksquare$ \textbf{\textit{Takeaway.}}} Singlepath load balancing at sub-flow granularity and source routing can achieve the same properties as packet spraying, while maintain the flow context for congestion control to react to failures in a timely manner. 

\section{Ethereal Transport for AI}

Reflecting on our observations in \S\ref{sec:limitations} and \S\ref{sec:single-multi}, we design a singlepath\footnote{We note that the notion of singlepath and multipath in our context is from a NIC's state and congestion control point-of-view.} transport, \name, suitable for distributed training workloads in CLOS-based topologies. We consciously design our algorithm to be simple, distributed, and efficient, requiring minimal changes to the NIC hardware.

\name has three main components \first flow interceptor  \second  path assignment \third loss recovery and handling failures. Algorithm~\ref{alg:ethereal} outlines \name's logic. The crux of our design relies on Theorem~\ref{th:equiv}, to achieve near-optimal load balancing using singlepath transport. We next describe each of these components in detail.

\subsection{Flow Interceptor}

The communication library (such as NCCL) on each server is responsible for launching flows using send calls, which trigger a series of functions that invoke RDMA verbs. We introduce a flow interceptor right before a send call is issued to the NIC, serving two key purposes.  

First, the flow interceptor acts as an entry point for \name's load balancing, opportunistically batching consecutive send calls from the communication library (\eg in an all-to-all communication). This batching mechanism allows \name to assign paths to multiple flows simultaneously, reducing the overhead of flow splitting and path assignment.  

Second, the flow interceptor mitigates load imbalance caused by flow synchronization --- an issue commonly observed in GPU clusters. It achieves this by randomizing both the start time of each flow and its position in the active list of queue pairs. By staggering flow start times, \name desynchronizes transmissions, reducing the likelihood of repetitive incasts. Additionally, randomizing a flow's position in the queue pair list further prevents structured transmission patterns that could lead to congestion at specific links.

Finally, the flow interceptor extracts the flow sizes, sorts the batch of flows based on the destination ToR, and forwards them for path assignment. Any rack-local flows are transmitted immediately.

\setlength{\textfloatsep}{0.2cm}
\setlength{\floatsep}{0.2cm}
\begin{algorithm}[!t]
\small 
	\SetKwFunction{arrival}{\textbf{\textsc{\textcolor{myred}{flowArrival}}}}
	\SetKwFunction{ack}{\textbf{\textsc{\textcolor{myred}{Ack}}}}
	\SetKwFunction{nack}{\textbf{\textsc{\textcolor{myred}{NAck}}}}
	\SetKwFunction{flowtimeout}{\textbf{\textsc{\textcolor{myred}{flowTimeout}}}}
	\SetKwFunction{receive}{\textbf{\textsc{\textcolor{myred}{receive}}}}
	\SetKwFunction{selectpath}{\textbf{\textsc{\textcolor{myred}{selectPath}}}}

	\SetKwProg{Fn}{function}{:}{}
	\SetKwProg{Proc}{procedure}{:}{}
	\SetKwInOut{KwIn}{Input}
	\SetKwInOut{KwOut}{Output}

	\KwIn{\ Batch of flows $\mathcal{F}$ ($n_{i,j}\ge 0$ towards each leaf) \\ \ Number of uplinks $s$}

	\Fn{\selectpath{$\mathcal{F}$}}{
		\hspace*{-\fboxsep}\colorbox{alg2}{\parbox{0.8\linewidth}{%
				\Comment{\textcolor{darkgray}{\textit{Uniform load-balancing}}}
				\vspace{1mm}
				
				\For{each leaf}{

					assign $\lfloor \frac{n_{i,j}}{s}\rfloor$ flows to each uplink

					$r = $ remaining flows

					$g = \mathrm{gcd}(r,s)$

					split each remaining flow to $\frac{s}{g}$ flows

					\For{each uplink}{
						assign $\frac{r}{g}$ remaining flows
					}

				}

			}}
	}

	\Fn{\arrival{$\mathcal{F}$}}{

		\hspace*{-\fboxsep}\colorbox{alg1}{\parbox{0.8\linewidth}{%
		\Comment{\textcolor{darkgray}{\textit{Mitigates repetitive incasts}}}
			\vspace{1mm}

			\For{each flow $\in \mathcal{F}$ }{
				Schedule after small random interval:
				
				\Indp
				
				$f\rightarrow$ nextAvail = now

				Insert $f$ to a random position in the list of queue pairs
			}

		}}
	}

	\Fn{\ack{}}{
		DcTcp() \Comment{\textit{Any CCA could be used}}
	}

	\tcc{----------------------------------------------------------------------------------------------------------}
	\tcc{\textcolor{blue}{Rare scenarios}}

	\Fn{\nack{}}{

		Select new path \Comment{\textit{Reroute}}

		GoBackN()	\Comment{\textit{Loss recovery}}
	}

	\Fn{\flowtimeout{$f$}}{

		\hspace*{-\fboxsep}\colorbox{alg3}{\parbox{0.8\linewidth}{%
			\Comment{\textcolor{darkgray}{\textit{Handles link failures}}}
			
			Select new path
			
			GoBackN()

			Notify failure

		}}
	
	}
	\Fn{\receive{}}{ 
		\hspace*{-\fboxsep}\colorbox{alg4}{\parbox{0.8\linewidth}{%

		\Comment{\textit{At the receiver}}

				\vspace{1mm}
				Extract path ID from the packet header

				Bit-flip and copy path ID to the acknowledgment

				sendAck()
		}}
	}

	\caption{\name}
	\label{alg:ethereal}
\end{algorithm}

\subsection{Divide and Conquer Load Balancing}
\label{sec:divide-conquer}

\name's load balancing algorithm opportunistically splits flows into multiple sub-flows and greedily distributes flows across the network topology to achieve near-optimal load balancing. The simplicity and effectiveness of \name owes it to the following assumptions:

\begin{itemize}[label=\small{\textcolor{myred}{$\blacksquare$}}]
	\item \textbf{Assumption 1:} The topology is CLOS-based and the entire topology structure, placement of GPUs (rank IDs), and link capacities are given as input to \name at initialization time.
	
	\item \textbf{Assumption 2:} The communication library (such as NCCL) is capable of handling data splitting and assembly upon arrival at the receiver. For instance, NCCL provides various configurations to split a message into multiple chunks and reassemble them at the receiver.
	
	\item \textbf{Assumption 3:} The NIC serves all the active queue pairs uniformly (e.g., round-robin) and does not enforce scheduling policies such as shortest queue first, which could potentially lead to load imbalance.
\end{itemize}

\noindent
\textbf{Path assignment to achieve uniform load balancing:} The \texttt{pathSelect} function in Algorithm~\ref{alg:ethereal} presents the pseudocode. \name's load balancing follows a greedy approach, \ie it assigns a batch of intercepted flows towards each ToR across the set of uplinks such that the overall congestion on both uplinks and downlinks closely approximates that of packet spraying.  

Specifically, let \name intercept $n_{i,j}$ flows on a server $i$ destined for a leaf switch $j$. Let $s$ denote the number of spine switches in a leaf-spine topology (or the number of core switches in a fat-tree topology). The algorithm first assigns $\lfloor\frac{n_{i,j}}{s}\rfloor$ flows to each spine switch, ensuring an even initial distribution of network load across all available paths toward ToR $j$. Up to this point, \name’s path assignment aligns with Alibaba HPN's flow allocation strategy, which assigns flows to the least congested path~\cite{10.1145/3651890.3672265}.  

However, any remaining flows ($r = n_{i,j} \bmod s$) pose a challenge, as the total number of flows is not evenly divisible by the number of paths. Simply assigning these remaining flows to the least congested path or relying solely on re-routing strategies can lead to path flapping, as discussed in \S\ref{sec:limitations}. Achieving near-optimal load balancing hinges on effectively handling these remaining flows.  

To address this, \name strategically \emph{splits only the remaining} $r = n_{i,j} \bmod s$ flows into $\frac{s}{g}$ subflows, where $g = \mathrm{gcd}(r,s)$. This transformation ensures that the number of flows becomes divisible by the number of paths. \name then assigns each of the $\frac{r}{g}$ flows to each spine switch, achieving a balanced distribution. From the proof of Theorem~\ref{th:equiv}, it follows that this path assignment results in congestion properties equivalent to packet spraying.

\myitem{Simple source routing to enforce the assigned paths:}
We enforce path assignment using a conceptually straightforward approach based on source routing. \name embeds path identifiers into packet headers, enabling switches to explicitly route packets based on these identifiers. This information is encoded in the source port field of the packet header. Specifically, the path identifier is represented as a $16$-bit value, where the first $8$ bits indicate the uplink index at the first-hop ToR switch, and the next $8$ bits denote the uplink index at the second-hop spine switch (if applicable for the packet's path).  

As a packet traverses the network, switches extract the path identifier and forward it to the corresponding uplink index. Each switch then swaps the first and last $8$ bits of the path identifier, simplifying the switching logic. This design ensures a fixed-function switching process, where each switch extracts only the first $8$ bits for forwarding and then swaps the bits before transmitting the packet to the next hop.  

Once a packet reaches the core, its path to the destination becomes unique. However, each switch along the downlink path continues performing the same bit-swapping operation. At the destination, the receiver NIC extracts the path identifier and copies it into the acknowledgment.  

Overall, this approach ensures that data packets and their corresponding acknowledgments follow the same set of links in the network, enabling unambiguous link failure detection, which we discuss next. Alternatively, \name could leverage approaches like RePac~\cite{273839} to achieve fine-grained control over path assignment without requiring modifications to switches. We leave it for future work to explore such approaches for source routing in \name.

\subsection{Loss Recovery and Handling Failures}
\label{sec:failures}

\myitem{Loss recovery solely based on vanilla RoCE:}  
To maintain compatibility with existing PFC-enabled RoCE implementations, \name does not modify loss recovery mechanisms and instead relies on NACK with Go-Back-N for handling packet drops in the network. Additionally, \name employs a per-flow timeout, which, upon expiry, triggers Go-Back-N and initiates loss recovery.  

\myitem{Reacting to failures with simplicity:}  
As described in \S\ref{sec:divide-conquer}, \name has knowledge of the topology at initialization. To this end, it maintains the state of each path in the network, referring to paths solely based on the $16$-bit encoding described earlier. Since \name operates at the application layer, this state information remains independent of the NIC.  

\name relies on two mechanisms to quickly detect failures (or slow links):  
\first The receiver immediately sends a NACK, triggering Go-Back-N and recovery at the sender.  
\second The sender detects failure when the per-flow timeout expires.  

Both mechanisms are standard in RoCE implementations on commodity hardware and assist in failure detection. However, while these mechanisms can detect failures, reacting to them in a timely manner requires additional support.  

Upon failure detection by either mechanism, the NIC assigns a new random path ID to the affected flow without redistributing all active flows. We chose this approach for its simplicity. However, a more sophisticated strategy could involve redistributing all active flows to achieve near-optimal load balancing. We leave this exploration for future work and currently prioritize a balance between simplicity and performance, specifically for flows on a failed path.

To incorporate the failed path into future load-balancing decisions, the NIC sends a control signal to \name via a separate completion queue (CQ). Upon receiving this signal, \name marks the path as ``bad'' and starts a timer to reset the path state. Subsequent flow arrivals undergo load balancing based only on the available ``good'' paths, ensuring near-optimal load balancing even under failures and resulting asymmetries.

\section{Evaluation}
\label{sec:evaluation}

We evaluate \name using Astra-Sim~\cite{won2023astrasim2,rashidi2020astrasim,astrasimweb} and MLCommons Chakra~\cite{chakra}. Our evaluation aims answering the following key questions:

\smallskip
\noindent
\textbf{Q1:} \textit{How does \name compare to packet spraying in terms of collective completion time?}

\name achieves collective completion times comparable to packet spraying while retaining the simplicity of singlepath transport. For large message sizes, \name reduces completion times by up to $37.99\%$ compared to packet spraying\footnote{Throughout this paper, we refer to end-host-based distributed packet spraying simply as packet spraying.}. Additionally, \name reduces the communication time per iteration of GPT-2 training by an average of $11.55\%$ across different collective algorithms in both leaf-spine and fat-tree topologies.  

\smallskip
\noindent
\textbf{Q2:} \textit{How does \name perform under network failures?}

\name reacts promptly to network failures, improving collective completion times by an average of $68.91\%$ compared to packet spraying and by $58.4\%$ compared to REPS.

\smallskip
\noindent
\textbf{Q3:} \textit{What is the overhead of \name's flow splitting in terms of NIC resources?}

Our evaluations show that \name incurs minimal overhead in terms of NIC resources. The increase in the number of queue pairs remains well within the limits supported by modern NICs. Specifically, \name uses at most $32$ queue pairs when employing recursive doubling in both leaf-spine and fat-tree topologies.

\begin{figure*}
\centering
\begin{subfigure}{0.48\linewidth}
\centering
\includegraphics[width=1\linewidth]{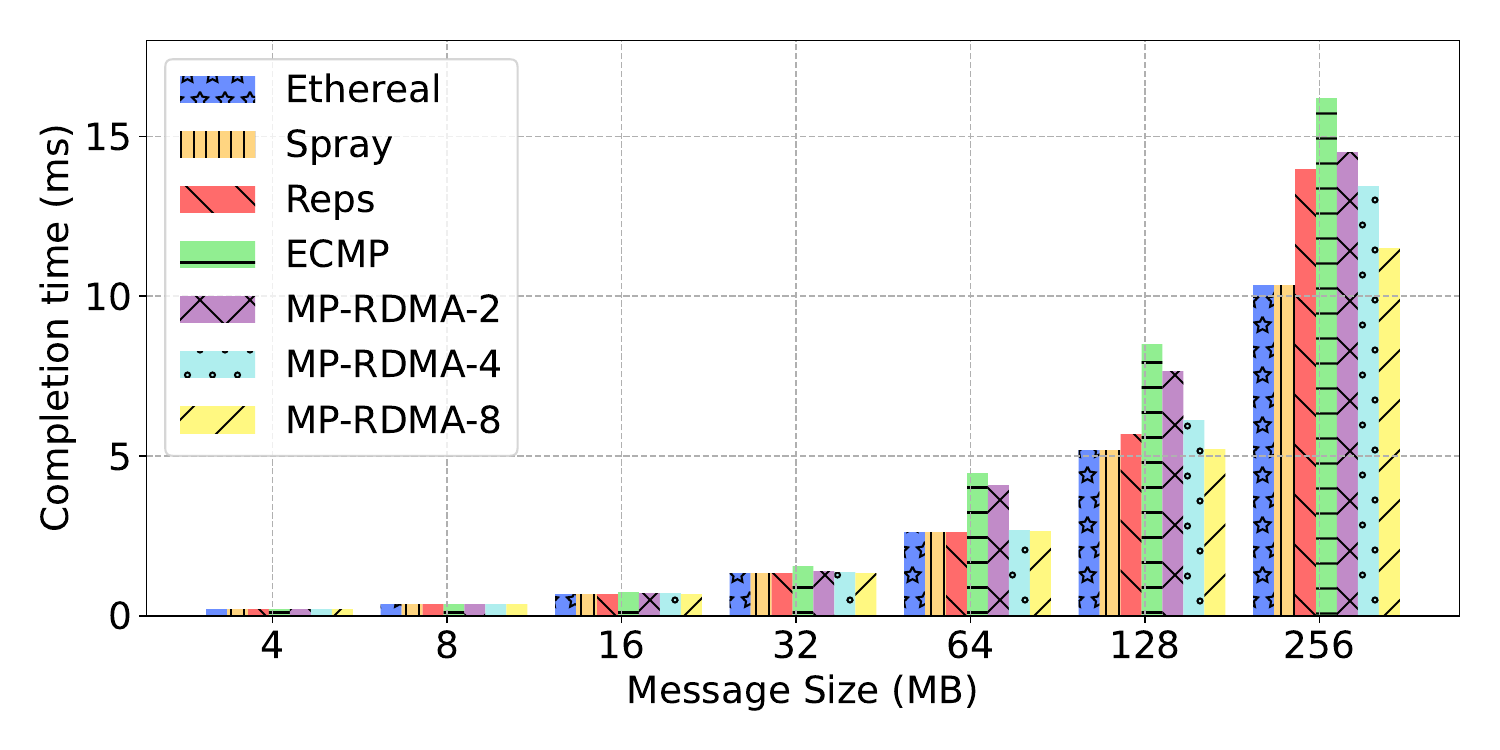}
\caption{All-to-All}
\label{fig:leaf-spine-a2a-cct}
\end{subfigure}
\begin{subfigure}{0.48\linewidth}
\centering
\includegraphics[width=1\linewidth]{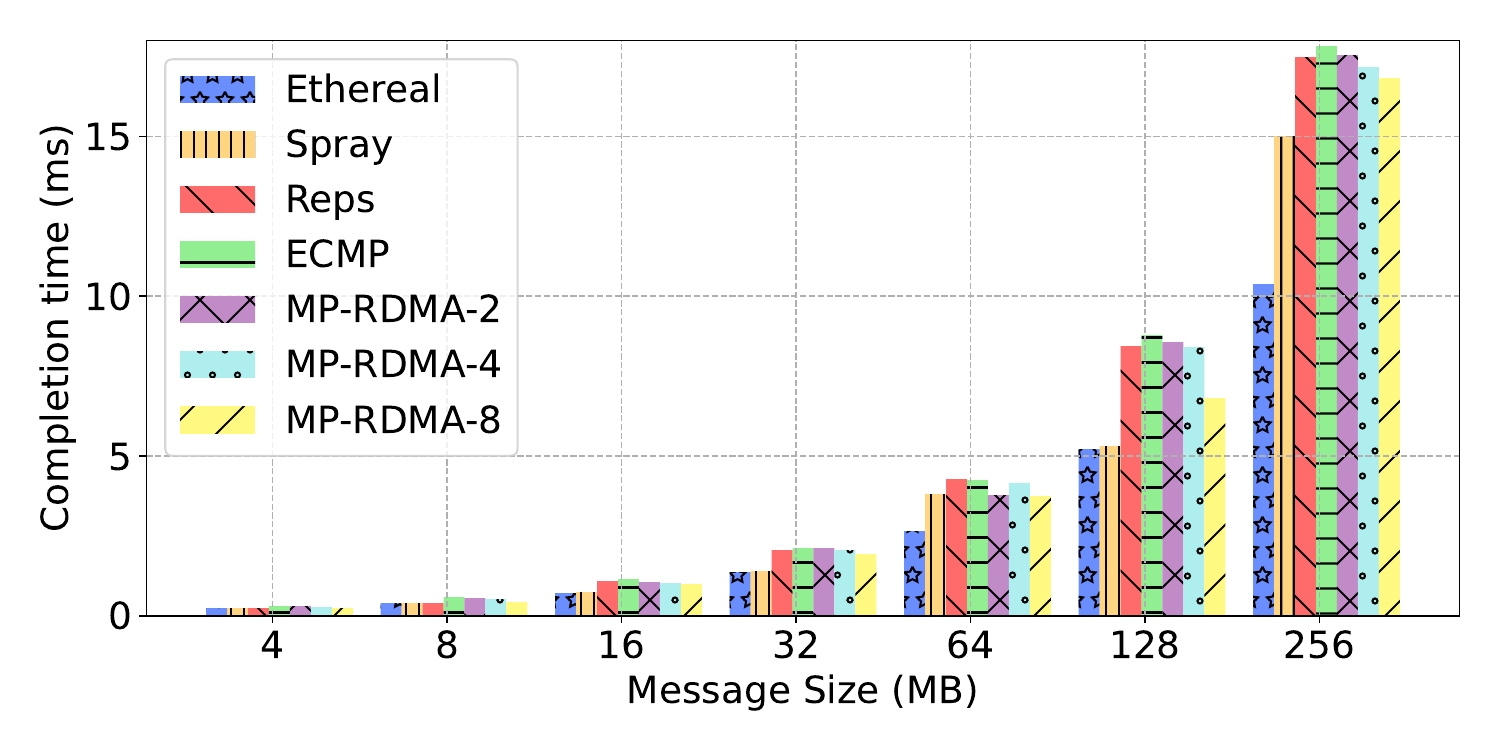}
\caption{Recursive Doubling}
\label{fig:leaf-spine-rd-cct}
\end{subfigure}
\begin{subfigure}{0.48\linewidth}
\centering
\includegraphics[width=1\linewidth]{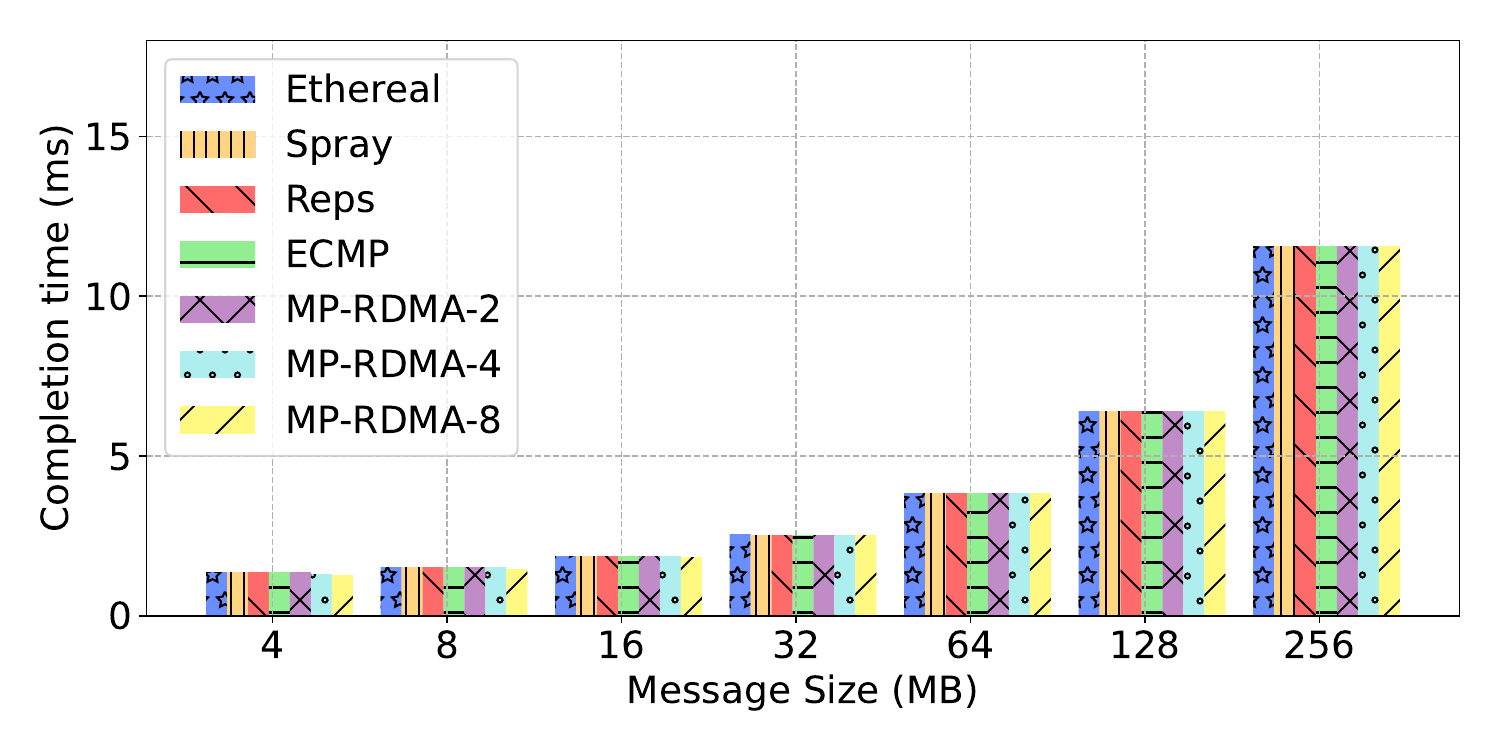}
\caption{Ring}
\label{fig:leaf-spine-ring-cct}
\end{subfigure}
\begin{subfigure}{0.48\linewidth}
\centering
\includegraphics[width=1\linewidth]{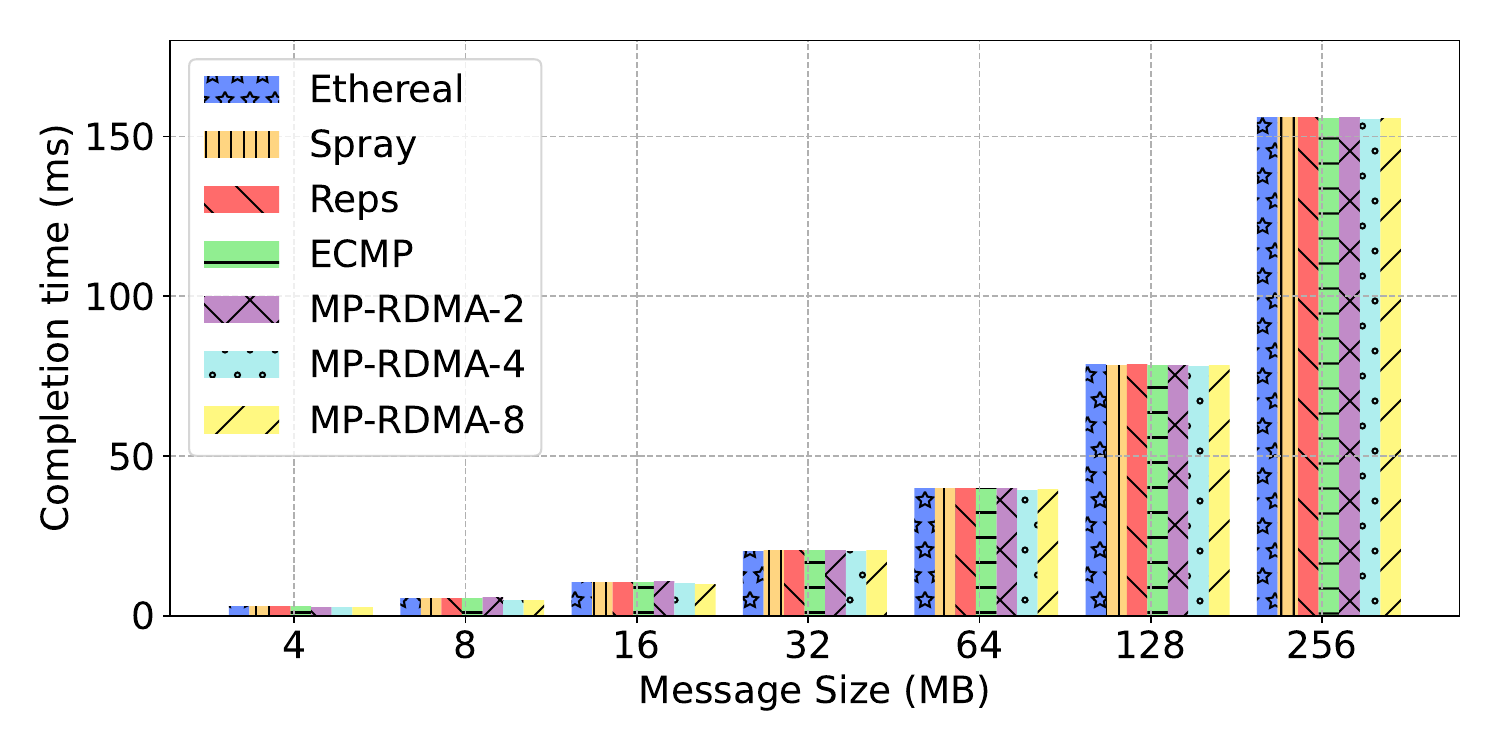}
\caption{Double Binary Tree}
\label{fig:leaf-spine-dbt-cct}
\end{subfigure}
\begin{subfigure}{0.24\linewidth}
\centering
\includegraphics[width=1\linewidth]{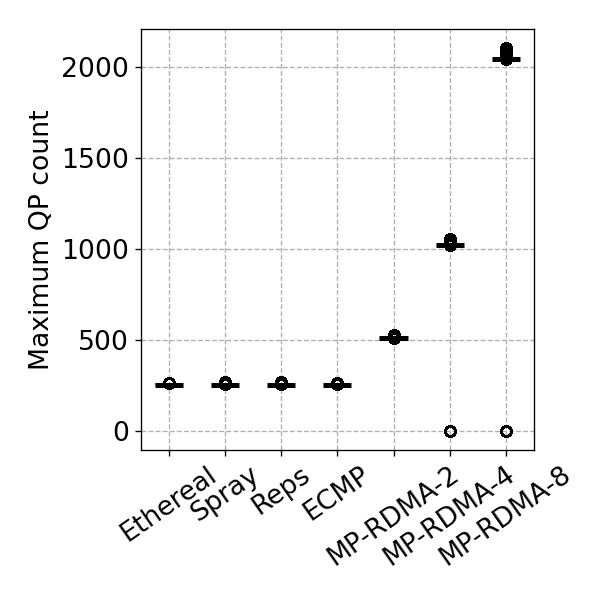}
\caption{Direct}
\label{fig:leaf-spine-a2a-qps}
\end{subfigure}\hfill
\begin{subfigure}{0.24\linewidth}
\centering
\includegraphics[width=1\linewidth]{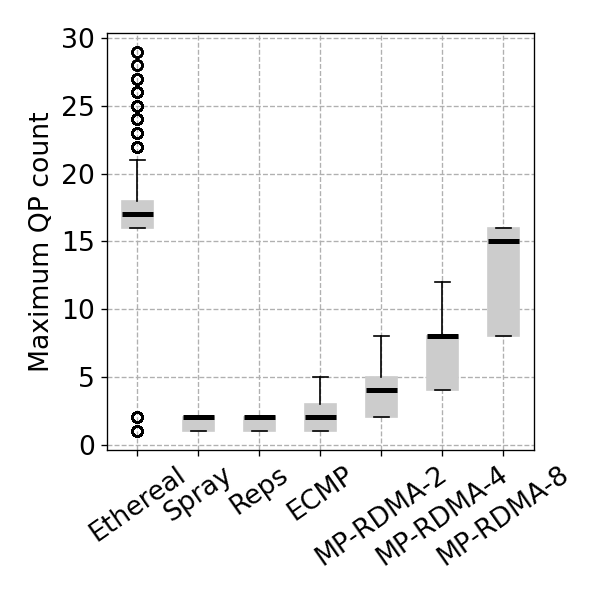}
\caption{Recursive Doubling}
\label{fig:leaf-spine-rd-qps}
\end{subfigure}\hfill
\begin{subfigure}{0.24\linewidth}
\centering
\includegraphics[width=1\linewidth]{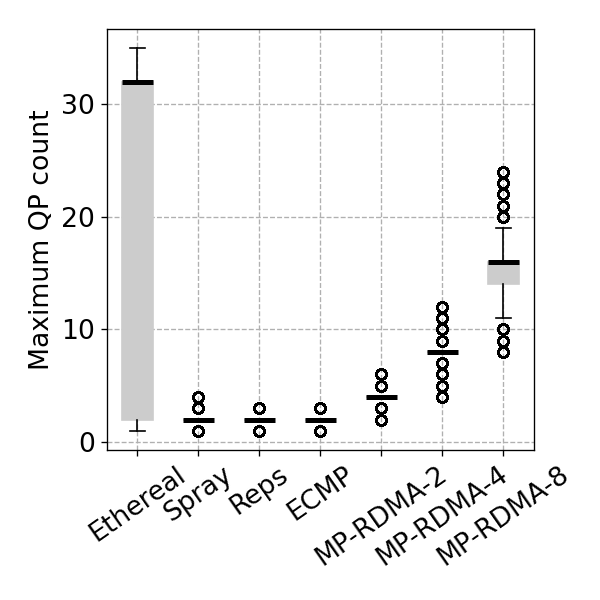}
\caption{Ring}
\label{fig:leaf-spine-ring-qps}
\end{subfigure}\hfill
\begin{subfigure}{0.24\linewidth}
\centering
\includegraphics[width=1\linewidth]{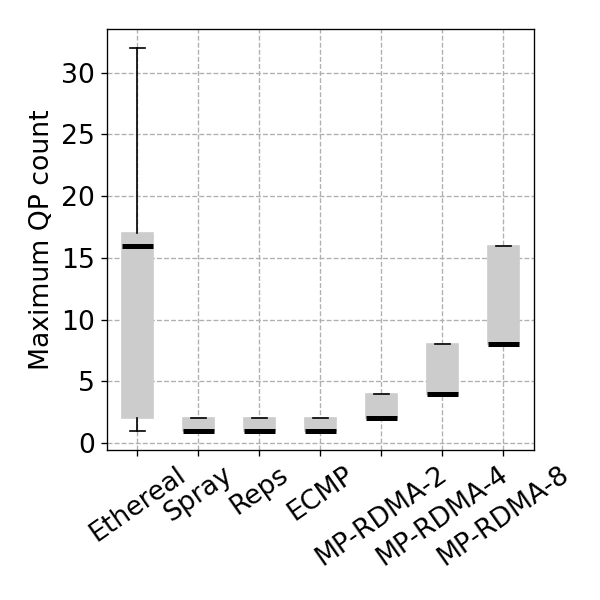}
\caption{Double Binary Tree}
\label{fig:leaf-spine-dbt-qps}
\end{subfigure}\hfill
\vspace{-2mm}
\caption{Leaf spine topology with $256$ GPUs: \name outperforms in completion times for all-to-all and recursive doubling without significantly increasing the number of required queue pairs at the NIC.}
\vspace{-2mm}
\label{fig:leaf-spine-collectives}
\end{figure*}

\subsection{Setup}

\myitem{Network Topologies:} We evaluate \name in leaf-spine and fat-tree topologies, two of the widely adopted topologies in practice~\cite{10.1145/3651890.3672265,10.1145/3651890.3672233}. The leaf-spine topology consists of $256$ GPUs, each with a dedicated NIC, arranged into $16$ ToR switches and $16$ spine switches. 
The fat-tree topology consists of $512$ GPUs with dedicated NICs, arranged into $32$ ToR switches, $32$ spine switches, and $16$ core switches. Each ToR switch is connected to $16$ NICs.
In both the topologies, each switch has $32\times 400$Gbps ports with a shared buffer of size $64$MB, similar to Broadcom Tomahawk $4$. In the fat-tree topology, switch to switch connections involve $4\times 400$Gbps ports, whereas in the leaf-spine topology, every switch to switch connection involves a single $400$Gbps port. We set the NIC bandwidth to $400$Gbps and use a $500$ns link latency (corresponding to $100$m fiber).

\myitem{Workloads:} We evaluate \name in two types of workloads:  
\first A single all-reduce operation using the all-to-all, recursive doubling, ring, and double binary tree collective communication algorithms, across message sizes ranging from $4$MB to $256$MB.  
\second One iteration of GPT-2 training under hybrid data-model parallelism, with a global batch size of $1024$ in the leaf-spine topology and $2048$ in the fat-tree topology, using different collective communication algorithms and a local batch size of $4$ per GPU.

\myitem{Baselines:} We compare \name against four load balancing approaches currently considered by the industry for large-scale GPU clusters:  
\first Packet spraying.  
\second REPS~\cite{reps2024,bonato2024smartt}.  
\third ECMP.  
\fourth MP-RDMA-$x$, which splits every flow by a factor of $x$.  
Together, these baselines span the spectrum of design choices in load balancing (Table~\ref{tab:design-choices}). We use DCTCP as the congestion control algorithm for all baselines and \name. Since packet spraying and REPS utilize multipath transport, we additionally employ per-packet timers for recovery and reorder buffers at the receiver. PFC is enabled for all baselines and \name.  
We set the $\alpha$ parameter for PFC dynamic thresholds to $1$. The timeout value is set to $1$ms --- per-queue for \name, ECMP, and RDMA, and per-packet for packet spraying and REPS.

\myitem{Metrics:} We primarily report collective completion times (CCT) for all-reduce operations and communication time per iteration for GPT-2 training. We also report the number of queue pairs used by each load balancing algorithm.

\begin{figure*}
\centering
\begin{subfigure}{0.48\linewidth}
\centering
\includegraphics[width=1\linewidth]{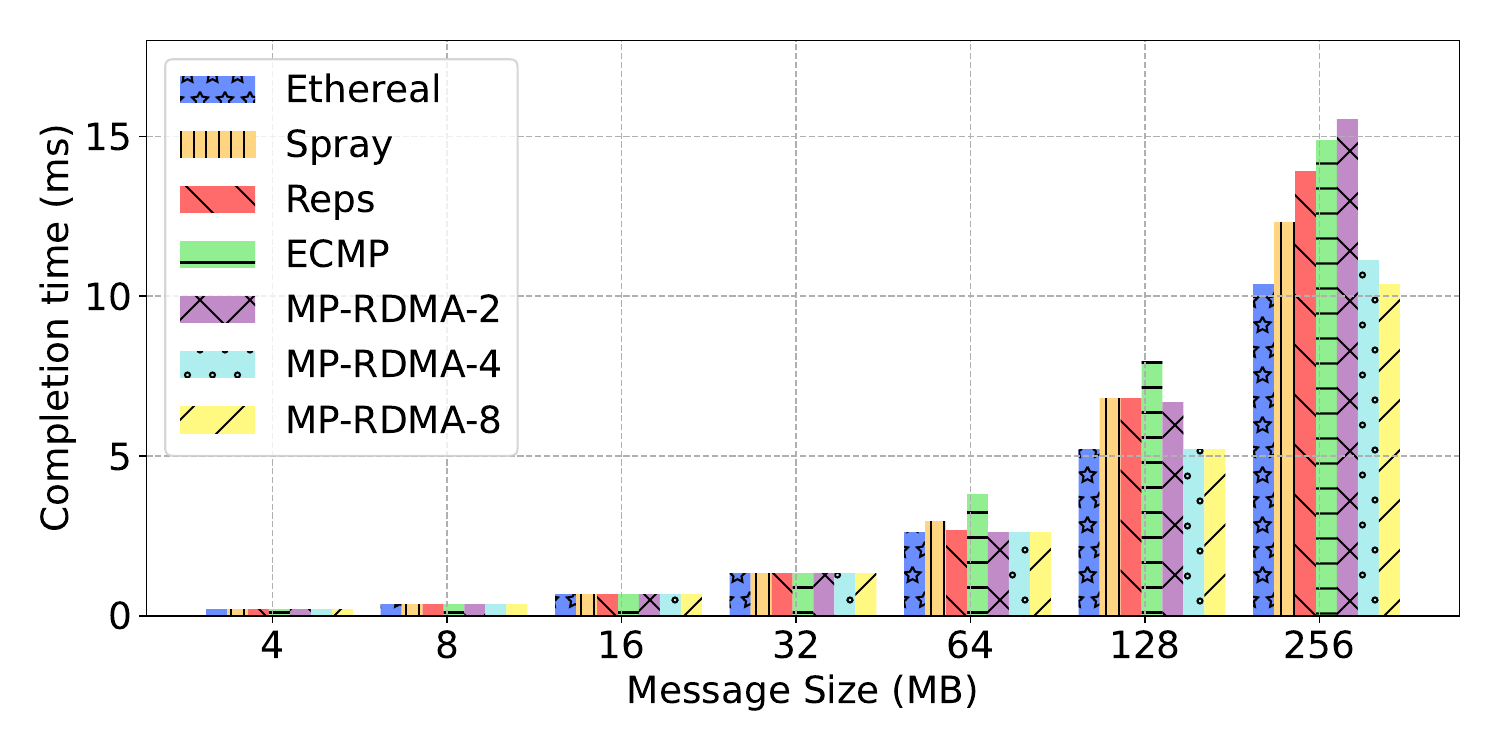}
\caption{All-to-All}
\label{fig:fat-tree-a2a-cct}
\end{subfigure}
\begin{subfigure}{0.48\linewidth}
\centering
\includegraphics[width=1\linewidth]{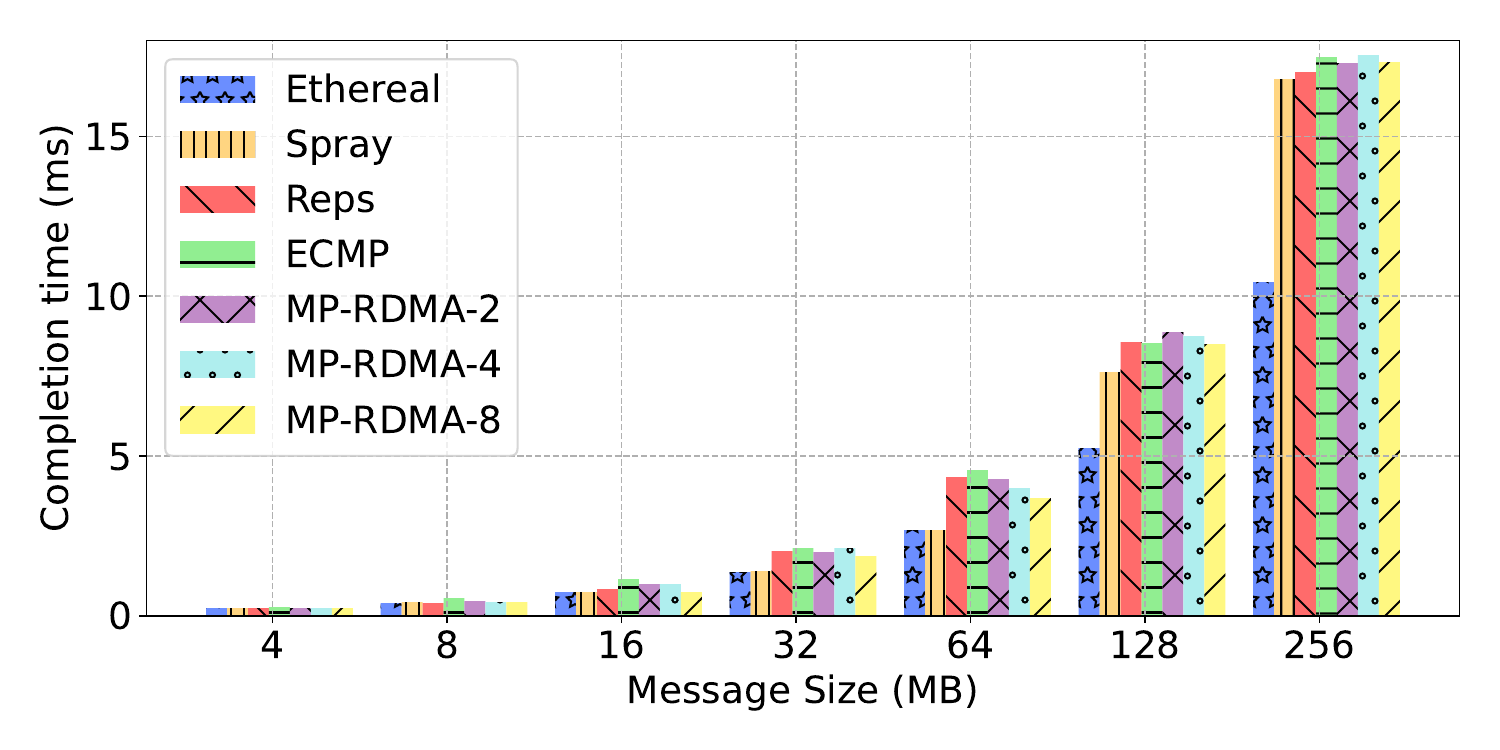}
\caption{Recursive Doubling}
\label{fig:fat-tree-rd-cct}
\end{subfigure}
\begin{subfigure}{0.48\linewidth}
\centering
\includegraphics[width=1\linewidth]{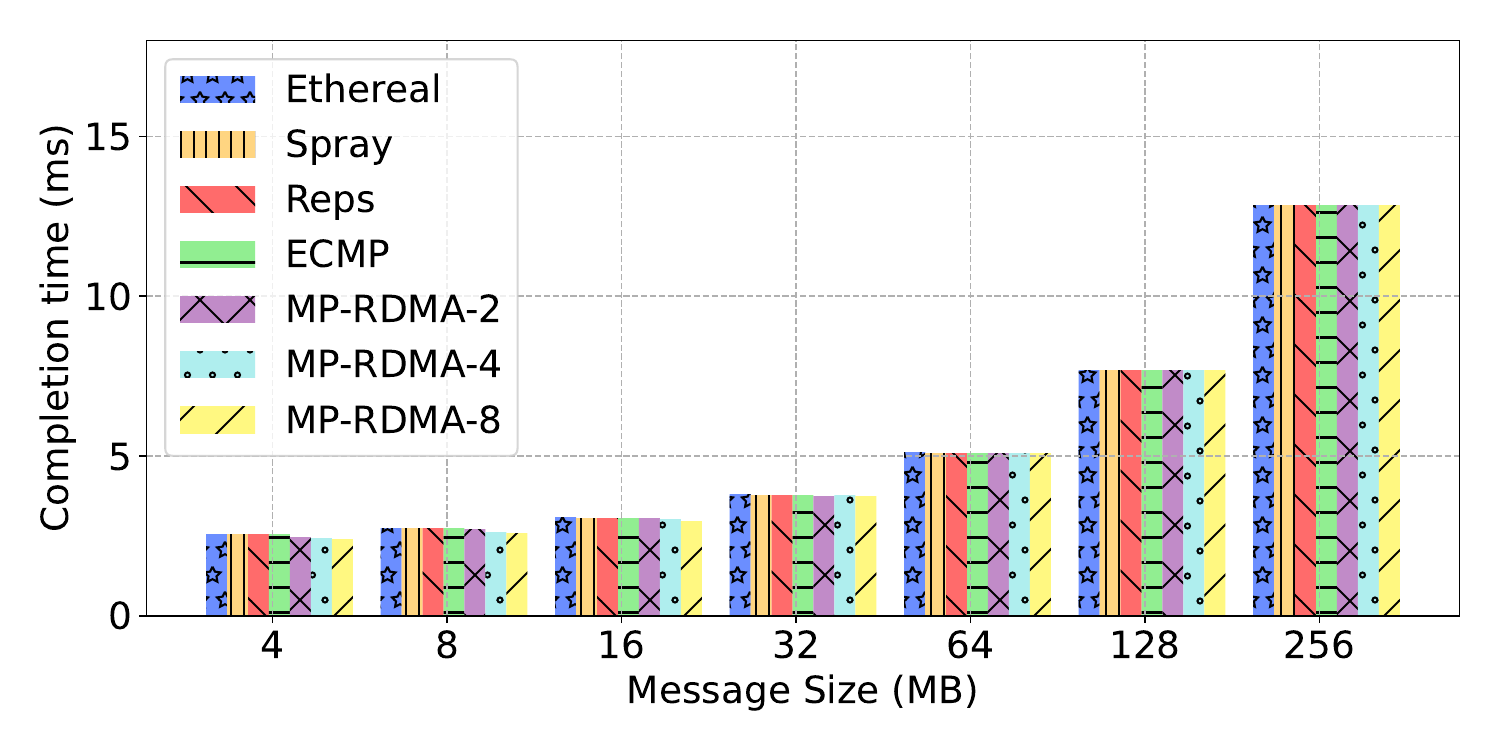}
\caption{Ring}
\label{fig:fat-tree-ring-cct}
\end{subfigure}
\begin{subfigure}{0.48\linewidth}
\centering
\includegraphics[width=1\linewidth]{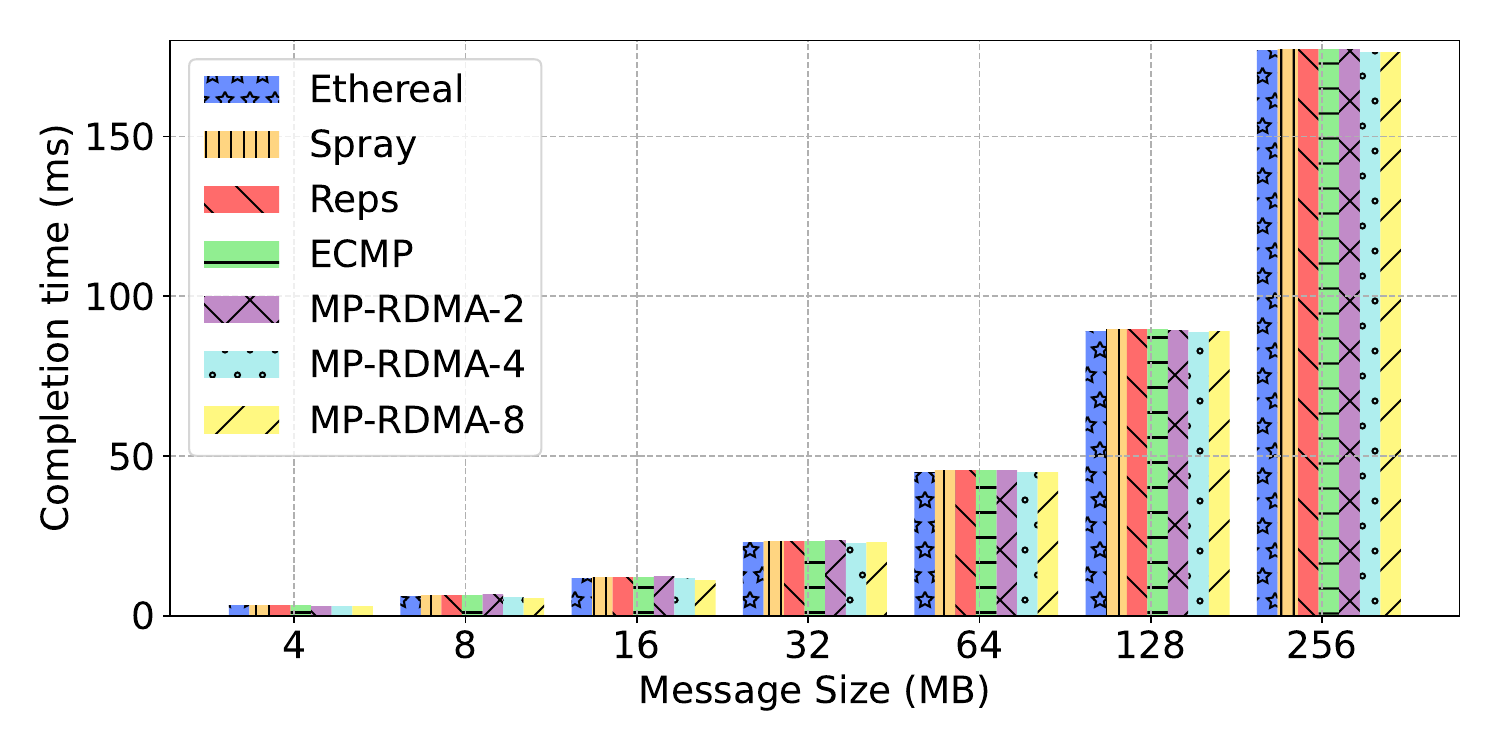}
\caption{Double Binary Tree}
\label{fig:fat-tree-dbt-cct}
\end{subfigure}
\begin{subfigure}{0.24\linewidth}
\centering
\includegraphics[width=1\linewidth]{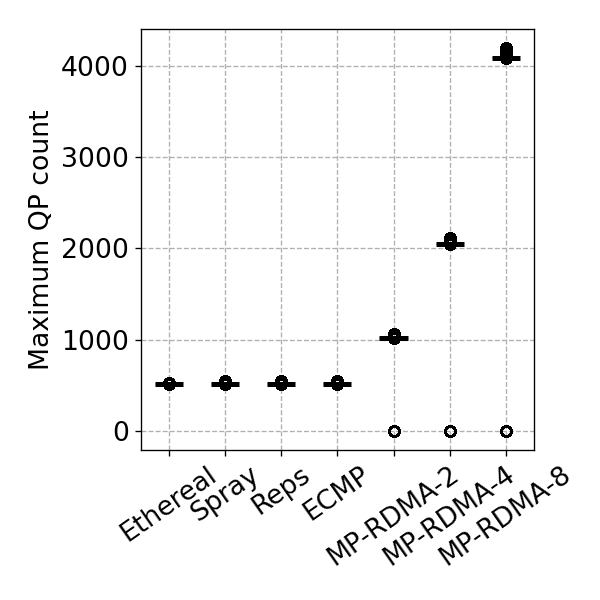}
\caption{Direct}
\label{fig:fat-tree-a2a-qps}
\end{subfigure}\hfill
\begin{subfigure}{0.24\linewidth}
\centering
\includegraphics[width=1\linewidth]{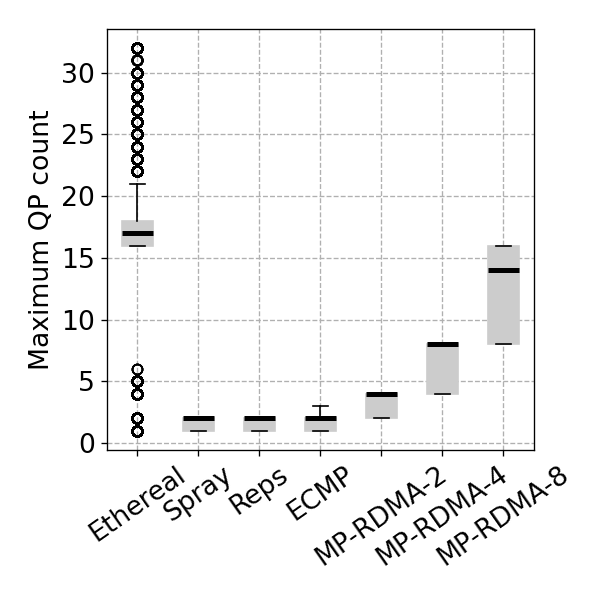}
\caption{Recursive Doubling}
\label{fig:fat-tree-rd-qps}
\end{subfigure}\hfill
\begin{subfigure}{0.24\linewidth}
\centering
\includegraphics[width=1\linewidth]{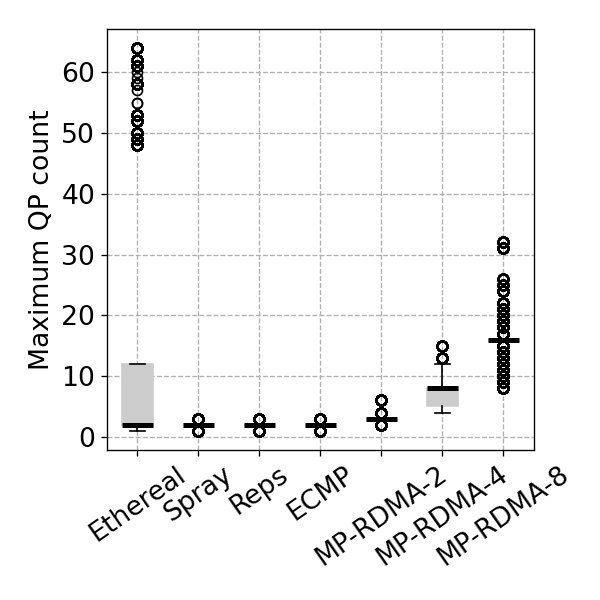}
\caption{Ring}
\label{fig:fat-tree-ring-qps}
\end{subfigure}\hfill
\begin{subfigure}{0.24\linewidth}
\centering
\includegraphics[width=1\linewidth]{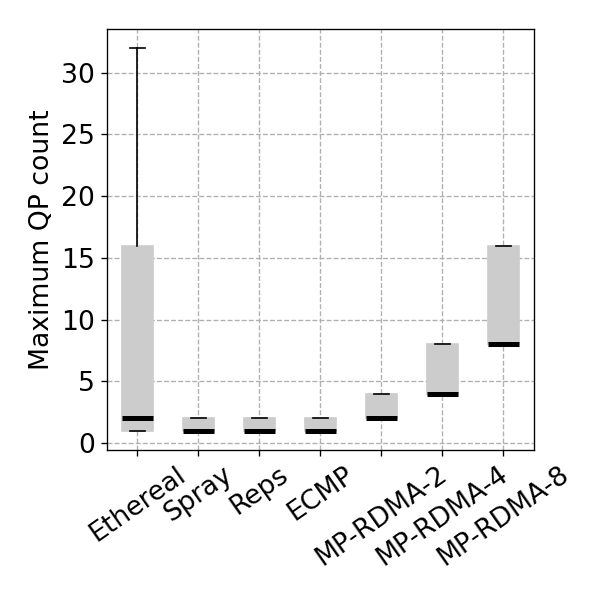}
\caption{Double Binary Tree}
\label{fig:fat-tree-dbt-qps}
\end{subfigure}\hfill
\vspace{-2mm}
\caption{Fat-tree topology with $512$ GPUs: Even in a hierarchical topology, \name significantly improves the completion times for allReduce with all-to-all and recursive doubling, while performing similarly to other approaches under ring and double binary tree.}
\vspace{-2mm}
\label{fig:fat-tree-collectives}
\end{figure*}

\subsection{Results}

\myitem{\name significantly outperforms in CCTs:}
Figure~\ref{fig:leaf-spine-collectives} shows the completion time for an allReduce operation using different collective algorithms in the leaf-spine topology. \name performs similar to alternative approaches for small message sizes, since there is little to no congestion in the network core. As the message size increases, \name starts to outperform ECMP even at $16$MB message size by $7.56\%$. While REPS performs close to \name for small message sizes, \name improves the completion times by $8.6\%$ at $128$MB and by $26.03\%$ at $256$MB message size for all-to-all communication; by $37.98\%$ and $40.65\%$ for $128$MB and $256$MB message sizes, respectively, for recursive doubling. 
\name performs similar to packet spraying for all-to-all communication and out-performs by $30.8\%$ for recursive doubling. MP-RDMA-$2,\ 4,\ 8$ generally show a monotonic improvement with the number of splits, but still lag behind \name in terms of completion times, particularly for larger message sizes. For ring and double binary tree, we observe similar performance under any load-balancing algorithm. This is mainly due to the limited communication neighbors in these algorithms, which do not require significant load balancing. Further, ring algorithm inherently has a uniform distribution of traffic over disjoint paths in a leaf-spine topology, leading to similar completion times across all load balancing algorithms.

We observe similar improvements with \name even in the fat-tree topology (Figure~\ref{fig:fat-tree-collectives}). For all-to-all with $128$MB message size, \name improves the completion times by $23.68\%$ compared to packet spraying, by $23.59\%$ compared to REPS, and by $34.81\%$ compared to ECMP. MP-RDMA improves over ECMP with more splittings and gets close to \name's performance. For recursive doubling with $256$MB message size, \name improves the completion times by $37.99\%$ compared to packet spraying, by $38.76\%$ compared to REPS, and by $40.41\%$ compared to ECMP. Interestingly, more splittings do not bring any significant improvements in completion times for MP-RDMA. This highlights that while flow splitting is essential to achieve uniform load balancing, careful path assignment is equally important. MP-RDMA splits flows as an attempt to increase the entropy to ECMP. However, the number of flows in recursive doubling is extremely low to have sufficient entropy to ECMP even by splitting \emph{every} flow. In contrast, \name splits only a tiny fraction of the flows and assigns paths explicitly to achieve near-optimal load balancing.

\myitem{\name reduces the time spent in communication:}
Figure~\ref{fig:leaf-spine-gpt} shows the time spent in communication over an iteration of GPT-2 training under different network load balancing algorithms in the leaf-spine topology. \name performs on-par compared to packet spraying with all-to-all direct collective algorithm used for allGather, ReduceScatter and allReduce operations during the iteration. Within this setup, \name improve the communication time by $15.59\%$ compared to REPS and by $27.51\%$ compared to ECMP. As expected, more splittings, show a clear monotonic improvement to communication times for MP-RDMA with $2$ splittings performing $23.5\%$ worse compared to \name, whereas $8$ splitting performing only $13.73\%$ worse. \name significantly outperforms other approaches in recursive doubling, reducing the communication time by $16.64\%$ compared to packet spraying, by $28.73\%$ compared to REPS and by $35.76\%$ compared to ECMP. 

We observe similar trends in the fat-tree topology (Figure~\ref{fig:fat-tree-gpt}). \name outperforms packet spraying and improves the communication time by $12.74\%$ with all-to-all and by $16.29\%$ with recursive doubling. With recursive doubling, \name improves the completion time by $30\%$ on average compared to the other baselines. Our results indicate how \name can significantly reduce the time spent in communication, making it an attractive option for distributed training workloads.

\begin{figure*}
\centering
\begin{subfigure}{0.24\linewidth}
\centering
\includegraphics[width=1\linewidth]{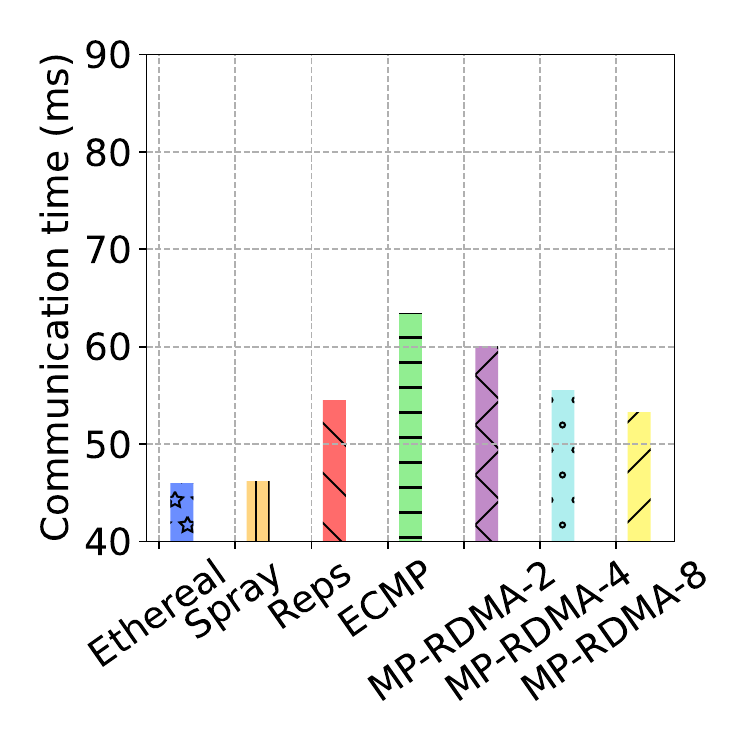}
\caption{All-to-All}
\label{fig:leaf-spine-a2a-cct-gpt}
\end{subfigure}
\begin{subfigure}{0.24\linewidth}
\centering
\includegraphics[width=1\linewidth]{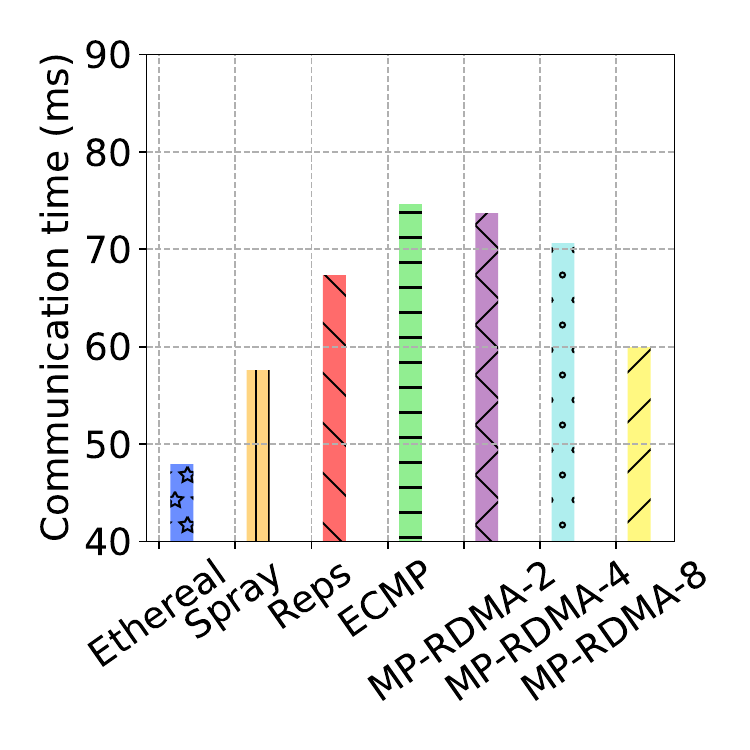}
\caption{Recursive Doubling}
\label{fig:leaf-spine-rd-cct-gpt}
\end{subfigure}
\begin{subfigure}{0.24\linewidth}
\centering
\includegraphics[width=1\linewidth]{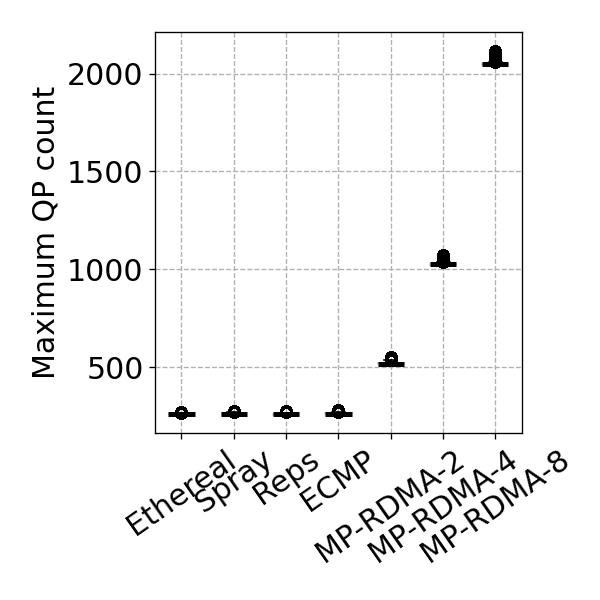}
\caption{Direct}
\label{fig:leaf-spine-a2a-gpt-qps}
\end{subfigure}\hfill
\begin{subfigure}{0.24\linewidth}
\centering
\includegraphics[width=1\linewidth]{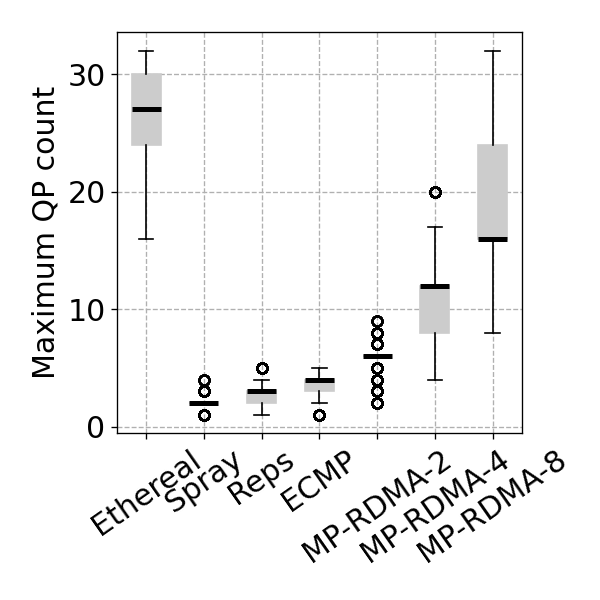}
\caption{Recursive Doubling}
\label{fig:leaf-spine-rd-gpt-qps}
\end{subfigure}\hfill
\vspace{-2mm}
\caption{Leaf spine topology GPT-2 (one iteration)}
\vspace{-2mm}
\label{fig:leaf-spine-gpt}
\end{figure*}

\begin{figure*}
\centering
\begin{subfigure}{0.24\linewidth}
\centering
\includegraphics[width=1\linewidth]{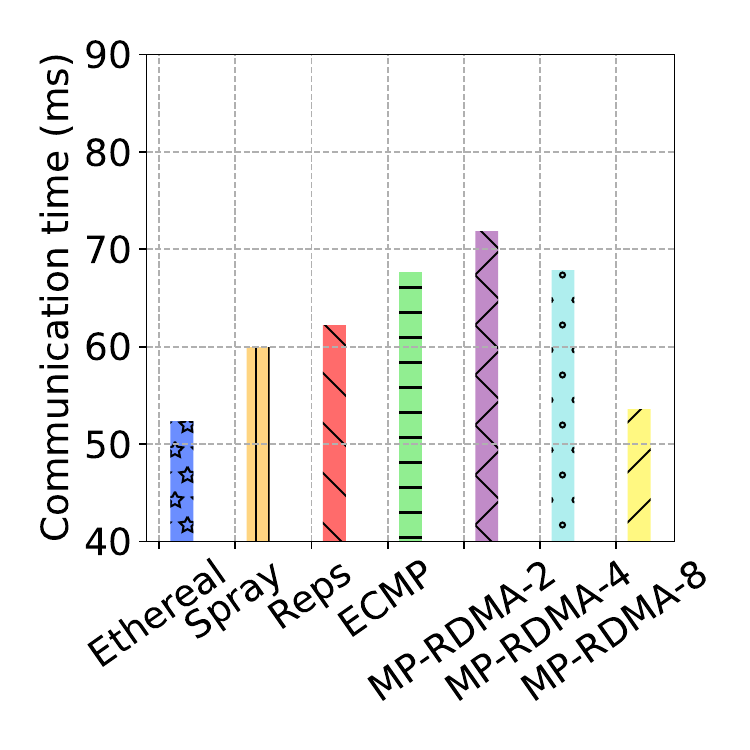}
\caption{All-to-All}
\label{fig:fat-tree-a2a-cct-gpt}
\end{subfigure}
\begin{subfigure}{0.24\linewidth}
\centering
\includegraphics[width=1\linewidth]{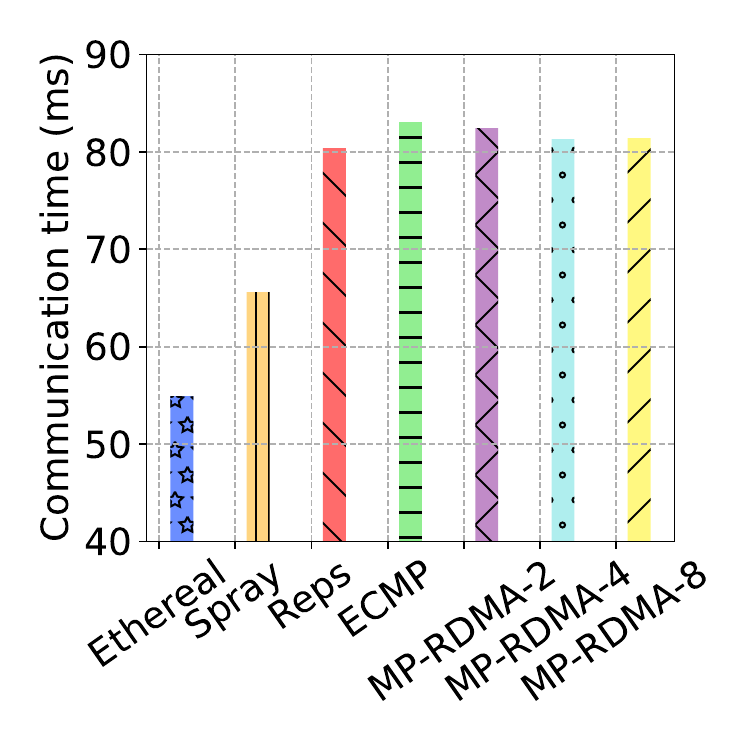}
\caption{Recursive Doubling}
\label{fig:fat-tree-rd-cct-gpt}
\end{subfigure}
\begin{subfigure}{0.24\linewidth}
\centering
\includegraphics[width=1\linewidth]{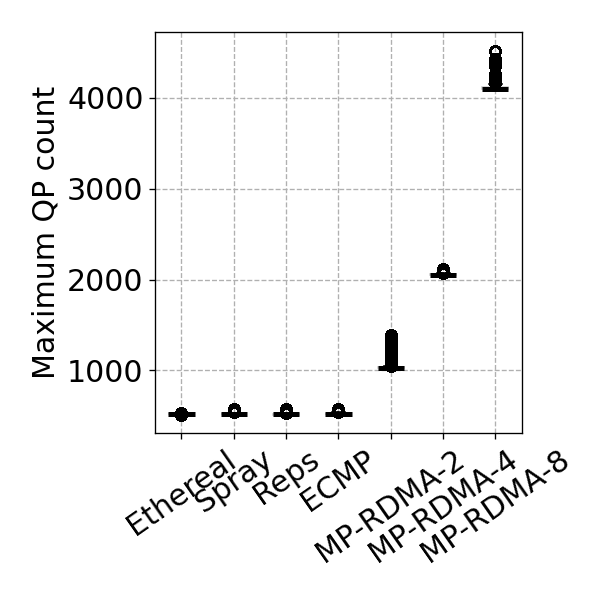}
\caption{Direct}
\label{fig:fat-tree-a2a-gpt-qps}
\end{subfigure}\hfill
\begin{subfigure}{0.24\linewidth}
\centering
\includegraphics[width=1\linewidth]{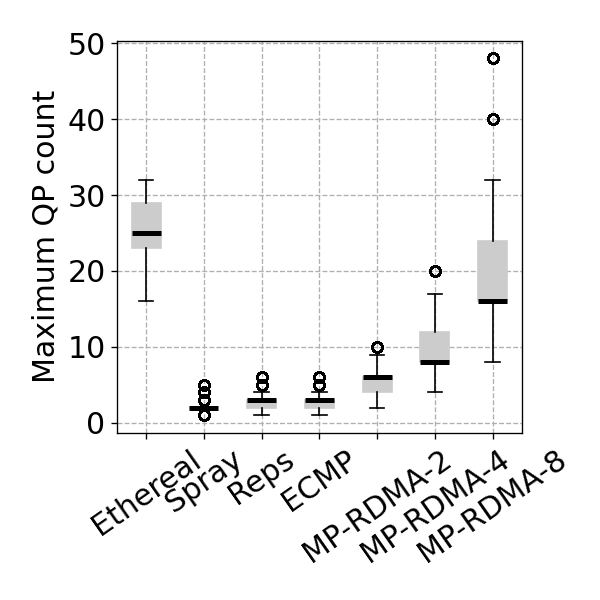}
\caption{Recursive Doubling}
\label{fig:fat-tree-rd-gpt-qps}
\end{subfigure}\hfill
\vspace{-2mm}
\caption{Fat Tree topology GPT-2 (one iteration)}
\vspace{-2mm}
\label{fig:fat-tree-gpt}
\end{figure*}

\myitem{\name does not incur significant overhead:}  
Given \name's superiority in completion times, we analyze its overhead, particularly in terms of additional NIC resource usage. Specifically, we measure this overhead as a function of the maximum number of queue pairs utilized by each load balancing algorithm. Figures~\ref{fig:leaf-spine-collectives},~\ref{fig:fat-tree-collectives},~\ref{fig:leaf-spine-gpt},~\ref{fig:fat-tree-gpt}, show the distribution of the number of queue pairs used by each algorithm. We observe that \name's QP utilization is well-known the capabilities of modern NICs. \name consumes at most $64$ queue pairs across all our evaluations, except for the all-to-all communication which inherently requires more queue pairs. With all-to-all, MP-RDMA-$8$ splits \emph{every} flow and maintains $>4000$ queue pairs, an order of magnitude higher than \name. This result highlights \name's ability to achieve near-optimal load balancing with minimal overhead in terms of NIC resources.

\myitem{\name outperforms even under link failures:}  
Figure~\ref{fig:failures} presents the completion times for the all-reduce operation under link failures in leaf-spine and fat-tree topologies. Specifically, after a collective starts, we fail a link between a ToR and a spine switch. The switch continues transmitting packets to the failed port for $100$ms before updating the routing table.  

\name's effectiveness in rapidly re-routing affected flows is evident, particularly for small message sizes. \name improves completion times by $66.3\%$ compared to packet spraying and by $56\%$ compared to REPS, on average, for message sizes between $4$MB and $64$MB. Although packet spraying employs per-packet timers for recovery, it continues to distribute packets toward the failed port, leading to excessive packet drops and increased completion times. REPS faces a similar issue: each new flow maintains its own cache of ``good'' paths and initially performs packet spraying in the first RTT for path exploration. Consequently, flows within a bandwidth-delay product (BDP) remain vulnerable to excessive drops and timeouts, as REPS continues transmitting toward the failed port.  

REPS underperforms even for flow sizes above BDP ($>64$MB), particularly in the fat-tree topology (Figure~\ref{fig:failures}). Since REPS explores paths in the first RTT and caches all entropies that do not experience congestion, it also caches the path that is about to fail. Once the link failure occurs, affected flows time out and reroute. These rerouted flows then cause congestion on the new path, triggering ECN marking and further rerouting of existing flows on that path, leading to \emph{path flapping}.  

\begin{figure*}
\centering
\begin{subfigure}{0.48\linewidth}
\centering
\includegraphics[width=1\linewidth]{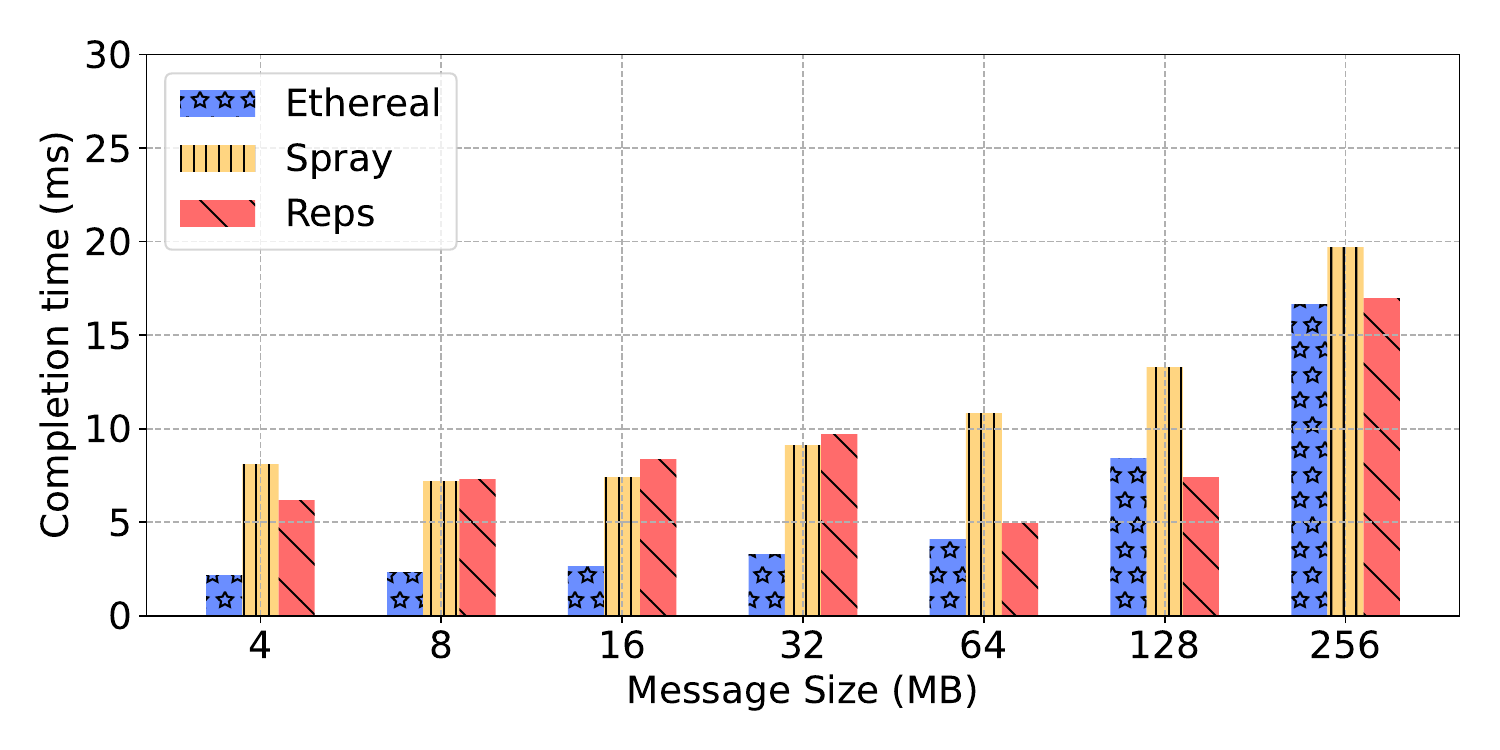}
\caption{All-to-All (leaf-spine)}
\label{fig:failures-leaf-spine-a2a-cct}
\end{subfigure}
\begin{subfigure}{0.48\linewidth}
\centering
\includegraphics[width=1\linewidth]{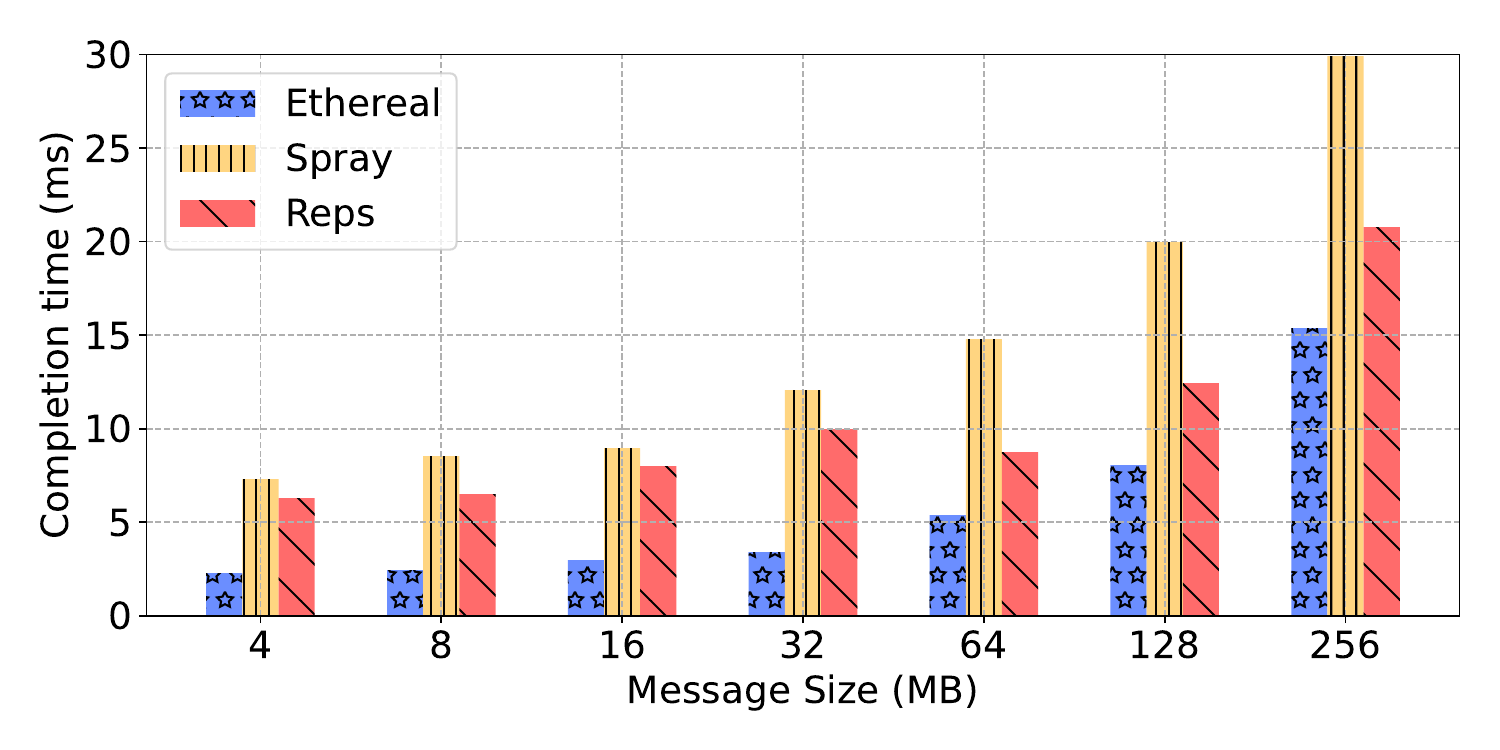}
\caption{Recursive Doubling (leaf-spine)}
\label{fig:failures-leaf-spine-rd-cct}
\end{subfigure}
\begin{subfigure}{0.48\linewidth}
\centering
\includegraphics[width=1\linewidth]{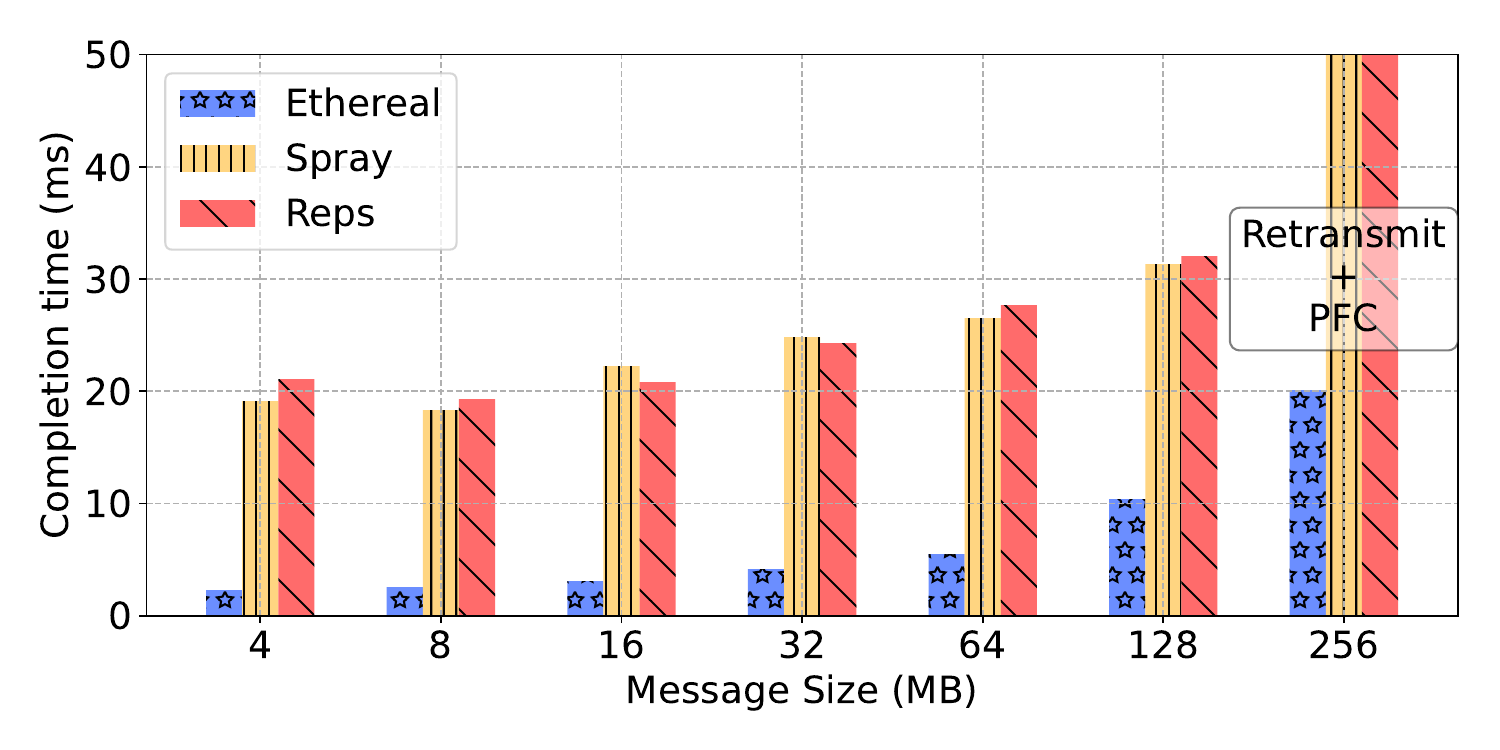}
\caption{All-to-All (fat-tree)}
\label{fig:failures-fat-tree-a2a-cct}
\end{subfigure}\hfill
\begin{subfigure}{0.48\linewidth}
\centering
\includegraphics[width=1\linewidth]{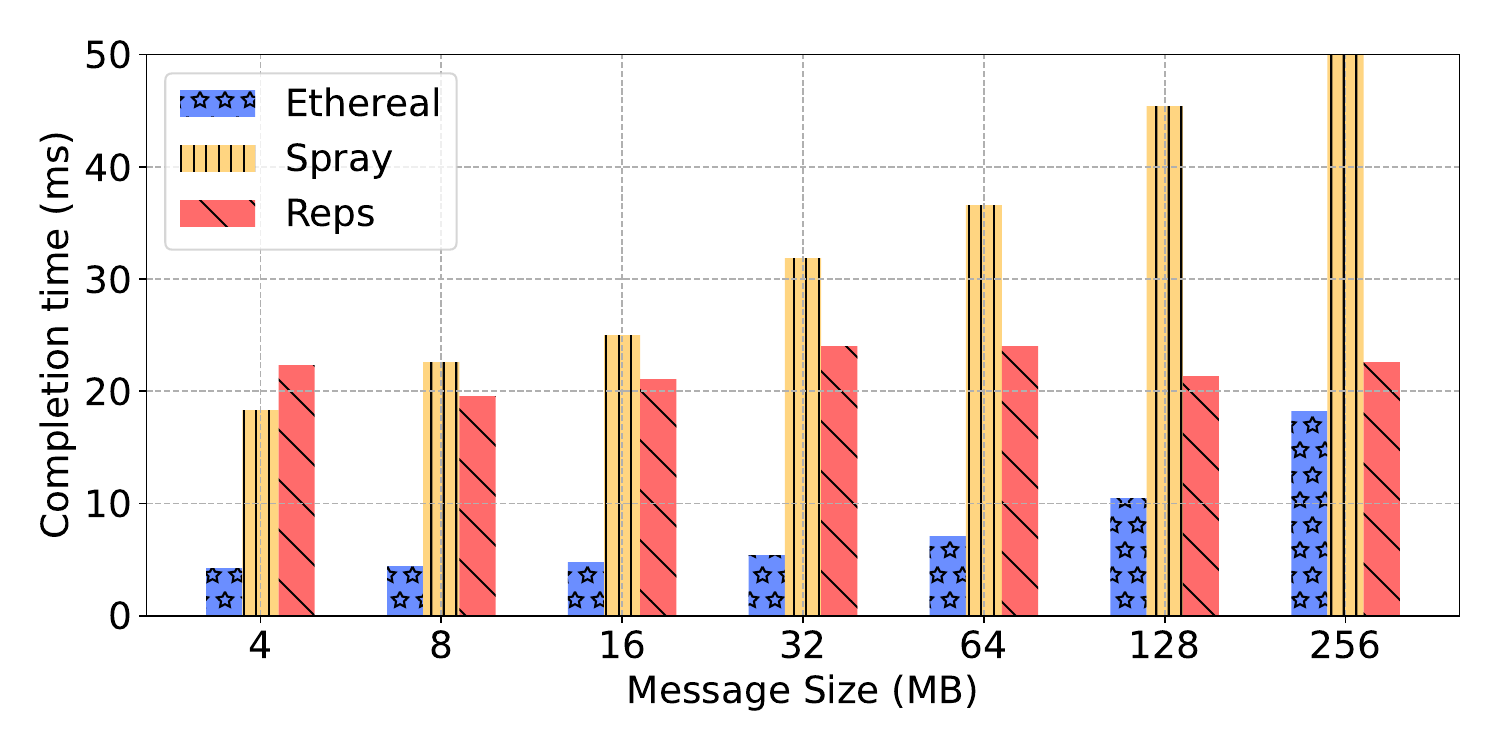}
\caption{Recursive Doubling (fat-tree)}
\label{fig:failures-fat-tree-rd-cct}
\end{subfigure}\hfill
\vspace{-2mm}
\caption{\name quickly re-routes flows upon detecting failures and significantly improves completion times compared to packet spraying and REPS.}
\vspace{-2mm}
\label{fig:failures}
\end{figure*}

Path flapping presents two major issues:  
\first It induces congestion across multiple paths, as discussed in \S\ref{sec:limitations}.  
\second Since flows initially transmit over paths that experienced no congestion in the first RTT, they eventually converge back onto the failed port, exacerbating packet drops. This behavior highlights the challenges adaptive routing strategies face during failures.  

In contrast, \name maintains per-host state for paths\footnote{Updates to the state of the paths are performed by affected flows individually, without requiring a central view (see \S\ref{sec:failures}).} and swiftly re-routes affected flows, resulting in improved completion times.

\section{Related Work}
\label{sec:relatedwork}

Over the past two decades, congestion control and load balancing have been an active area of research. We discuss the most relevant works in this section, particularly in the context of datacenter networks.
Several approaches to mitigate congestion in datacenters have been widely explored, with few deployed in production networks. ECN-based approaches such as DCTCP~\cite{10.1145/1851182.1851192} and DCQCN~\cite{10.1145/2785956.2787484} rely on ECN marking to detect congestion and adjust the sending rate accordingly. DCQCN is currently the default algorithm in RoCEv2 networks. Delay-based approaches such as TIMELY~\cite{10.1145/2785956.2787510} and SWIFT~\cite{10.1145/3387514.3406591} have been shown to be simple and effective in large-scale production networks. In-band network telemetry has emerged as an alternative to ECN, enabling fine-grained congestion control. HPCC~\cite{10.1145/3341302.3342085} reacts to inflight bytes, whereas PowerTCP~\cite{278346} reacts to both inflight bytes and queue gradients, by leveraging in-network telemetry. In terms of load balancing, numerous works have explored the design space~\cite{10.1145/3098822.3098839,259355,10.1145/2890955.2890968,10.1145/2619239.2626316}, ranging from flowlet switching to fine-grained per-packet load balancing, both at the switch and at the end-host. 
These approaches were primarily motivated by the unpredictable traffic patterns that are common in storage and search workloads. In the context of distributed training, the repetitive traffic patterns and the specific properties of the workload led to the design of algorithms such as MLTCP~\cite{10.1145/3696348.3696878}, CASSINI~\cite{295625}, gradient trimming~\cite{10.1145/3696348.3696880}, and MLT~\cite{295627}. While MTCP and CASSINI focus on congestion scenarios when multiple jobs compete, gradient trimming and MLT focus on distributed training in dedicated clusters. \name is aimed at the latter, keeping the design simple and efficient, while achieving near-optimal load balancing. Major efforts are currently being made to improve load balancing in GPU clusters, notably REPS~\cite{reps2024,bonato2024smartt}, the Ultra Ethernet Consortium~\cite{uec}, with several industry talks frequently being presented at OpenCompute Project (OCP)~\cite{ocp}. In comparison to the recent works, \name is the first to combine flow splitting and source routing to achieve near-optimal load balancing without requiring significant changes to the NIC hardware. 

\section{Conclusion}

This work challenges the prevailing belief that packet spraying is necessary to improve the performance of large-scale distributed training workloads. We presented \name, a simple alternative approach that relies solely on the existing singlepath transport protocol from a NIC's perspective. 
\name~does not require any changes to NIC hardware and can be implemented entirely in software.
We analytically demonstrated that optimal load balancing can be achieved by splitting a tiny fraction of the flows upon arrival. Our workload-driven empirical evaluations further show that \name~can significantly improve 
collective completion times and the time spent in communication. \name~is particularly useful in scenarios with link failures, since identifying and avoiding a failed path is relatively straightforward.

We hence believe that \name~offers an interesting alternative perspective for developing next-generation transport protocols tailored to large-scale distributed training, and that our results warrant further research on transport protocols.
In particular, it would be interesting to also explore how to exploit topological symmetries and characteristic workload properties also for inference workloads.

\section*{Acknowledgments}
This work is part of a project that has received funding
from the European Research Council (ERC) under the European Union’s Horizon 2020 research and innovation programme, consolidator project Self-Adjusting Networks (AdjustNet), grant agreement No. 864228, Horizon 2020, 20202025.
\begin{figure}[!h]
    \centering
    \includegraphics[width=0.5\linewidth]{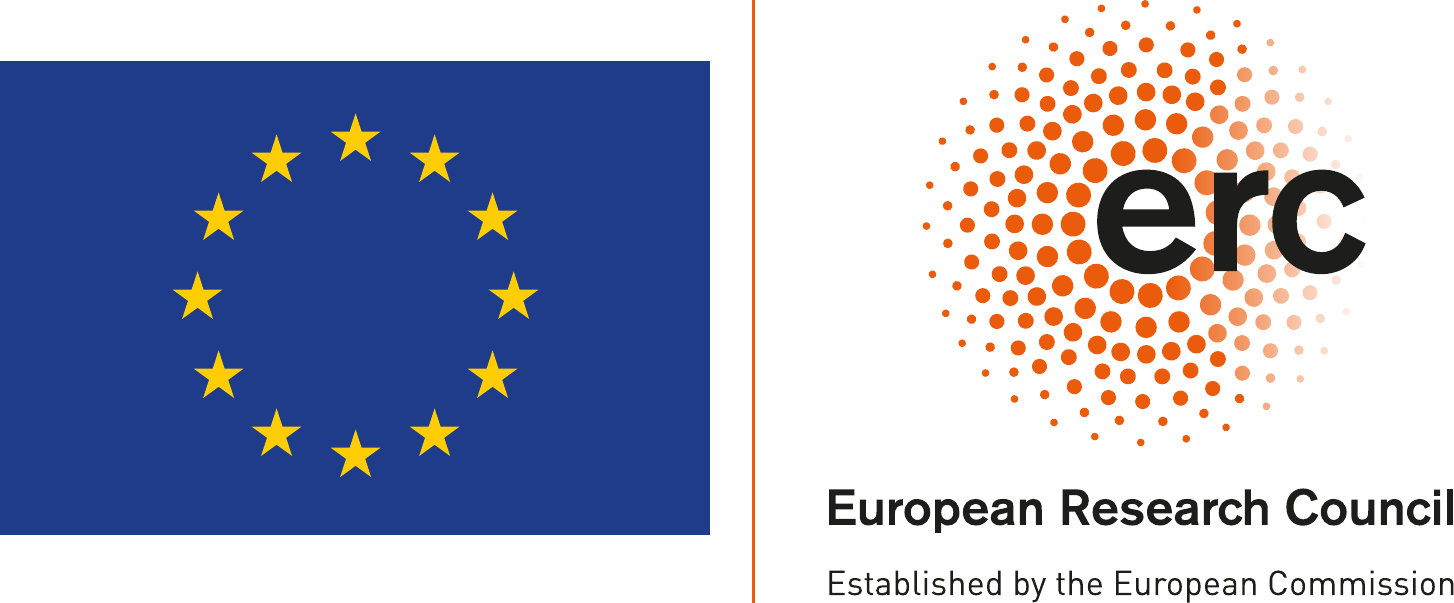}
    \label{fig:my_label}
\end{figure}

\label{bodyLastPage}

\bibliographystyle{plainurl}
\bibliography{references.bib}

\label{LastPage}
\end{document}